\documentclass[11pt,a4paper]{article}
 
\usepackage{amsmath,amsthm,amssymb}
\usepackage{graphics,graphicx}
\usepackage{epsfig}
\usepackage{multicol}
\usepackage{color}
\makeatletter
\@addtoreset{equation}{section}
\makeatother

\begin{document}
\begin{flushright}
{QMUL-PH-09-15 \\ UVIC-TH-09-09}\\
\end{flushright}
\begin{center}
\Large{\bf String Necklaces and Primordial Black Holes from Type IIB Strings}
\end{center}
\begin{center}
\large{\bf Matthew Lake,} ${}^{a, b,}$\footnote{m.lake@qmul.ac.uk} \large
{\bf Steven Thomas} ${}^{a,}$\footnote{s.thomas@qmul.ac.uk}  \large{\bf and John Ward} ${}^{c,}$\footnote{jwa@uvic.ca}
\end{center}
\begin{center}
\emph{ ${}^a$ Center for Research in String Theory, Queen Mary University of London \\ Mile End Road, London E1 4NS, UK \\}
\vspace{0.1cm}
\emph{ ${}^{b}$ Astronomy Unit, School of Mathematical Sciences, Queen Mary University of London \\ Mile End Road, London E1 4NS, UK \\}
\vspace{0.1cm}
\emph{${}^{c}$ Department of Physics and Astronomy, University of Victoria, Victoria, BC \\ V8P 1A1, Canada }
\end{center}
%%%%%%%%%%%%%%%%%%%%%%%%%%%%%%%%%%%
\begin{abstract}
We consider a model of static cosmic string loops in type IIB string theory, where the strings wrap cycles within the internal space. The strings are not
topologically stabilised, however the presence of a lifting potential traps the windings giving rise to kinky cycloops. 
We find that PBH formation occurs at early times in a small window, whilst at late times we observe the formation of dark matter relics in the scaling regime. 
This is in stark contrast to previous predictions based on field theoretic models. 
We also consider the PBH contribution to the mass density of the universe, and use the experimental data to impose bounds on
the string theory parameters.
\end{abstract}
%%%%%%%%%%%%%%%%%%%%%%%%%%%%%%%%%%%
\section{Introduction}
In recent years there has been a renewed effort to test string theory in a cosmological context. This is due in part to the availability of increasingly 
precise data from experiments such as WMAP \cite{Komatsu:2008hk} and SDSS \cite{AdelmanMcCarthy:2005se}. 
However, it is also due to theoretical advances which have resulted in a better understanding of the 
compactifications of the theory down to $(3+1)$-dimensions. Since such compactifications are typically warped this means that 
mass scales in the effective theory can be significantly reduced. One important consequence of this is that superstrings may have 
a much smaller tension than first realised \cite{Copeland:2003bj, Majumdar:2005qc}.

Originally Witten \cite{Witten:1985fp} ruled out the notion that $F$-strings could play the role of cosmic-strings because of their extremely high mass density. 
The un-warped string tension is so large that the presence of cosmic $F$-strings (formed after inflation) would be immediately evident in the CMB. 
This effectively killed the field until GKP (Giddings-Kachru-Polchinski) \cite{Giddings:2001yu} showed, in the context of type IIB strings, 
that by turning on non-trivial fluxes threading cycles in a class
of compact manifolds one could obtain highly warped four-dimensional backgrounds. The warping acts in such a way as to reduce the overall tension
of any object, and therefore it was possible to evade the observational bounds.

Simultaneously there has also been renewed interest in models of open string inflation. 
In such models the energy density of the inflaton field(s) is provided by the geometric distance between the $D3$-brane of our 
universe and parallel branes or anti-branes.  Since such branes/anti-branes are charged under massless $RR$-form 
fields it is expected that large numbers of $F$ and $D$-strings, as well as bound states of multiple strings, will be produced at the end of 
inflation \cite{Majumdar:2005qc, Sarangi:2002yt, Firouzjahi:2006vp}.  
Generically these objects are called $(p,q)$-strings, since they are typically a bound state 
consisting of $p$ $F$-strings and q $D$-strings \cite{Schwarz:1995dk, Thomas:2006ud, Firouzjahi:2006xa}. 
Observationally however this raises a potential problem as these strings are not red-shifted away and therefore (presumably) fine tuning is 
required to restrict their number to $\mathcal{O}(1)$ per Hubble radius at the present epoch\footnote{This is necessary for the 
string embedding considered here. We argue for a UV completion of cosmic string scenarios so we must consider the origin of the strings/necklaces in the parent theory. 
Since the most well modelled inflationary scenarios in type IIB string theory are brane inflation scenarios, we will take this as the inflationary mechanism.  
However brane inflation typically results in the formation of large numbers of defects after the inflationary epoch. 
Since we do not observe such objects in the visible horizon we must allow for some kind of tuning of the theory \emph{prior} 
to inflation which ensures that (in the post inflationary epoch) the relative numbers of branes, anti-branes and fluxes cancel 
leaving a small number density of residual defects.}. In this paper we will simply assume that some phase of open string inflation
has occurred, though it remains an open problem to generate such a configuration in an explicit
model of string theory inflation.

Given that these strings exist in a higher dimensional theory, one could also imagine that they wrap cycles within the internal space \cite{Avgoustidis:2004zt}. 
Such cycles could be either "smooth" or "lumpy" from a four dimensional perspective. 
A string that wraps a series of internal cycles at separate points in four-dimensional space would appear as a necklace. 
That is, as a system of monopoles or "beads" connected by string segments. Alternatively, strings which wrap the compact directions smoothly 
along their four dimensional length would appear to have a continuously varying mass density. The exact nature of the variation depends on the 
geometry of the compact space, though in general we would expect the effective tension to be periodic.

Matsuda has proposed that a necklace structure may form from a smoothly varying configuration as the string relaxes to a quasi-stable 
state \cite{Matsuda:2004bg}. In this case the periodic variation of effective string tension is viewed as the sum of the standard string tension 
plus a lifting potential in the (angular) compact directions. 
If the angular directions are flat the energy of the string is minimised for a smoothly varying configuration, resulting in a constant effective string tension, 
but in the presence of a potential a necklace structure is energetically favoured. 
Furthermore the existence, or absence, of a potential affects the stability of the extra-dimensional windings when strings 
chop off from the network to form loops. In the absence of a lifting potential windings must be stabilised topologically giving rise to objects 
called "cycloops". This is not true for necklace solutions which may exist, at least over cosmologically relevant time scales, 
even if the compact space is simply connected. The quasi-stability of the necklace solution is discussed in greater detail in section 5.5.

The cosmological consequences of cycloops were first investigated by Avgoustidis and Shellard \cite{Avgoustidis:2005vm}. 
They showed that, unlike an ordinary string loop, a decaying cycloop 
leaves a topologically trapped remnant when it reaches zero radius. This remnant appears as a monopole to a four dimensional observer and may be 
interpreted as a dark matter particle. Like ordinary string loops, cycloops also have a small probability ($f \sim 10^{-20}$) of collapsing to form PBH's in the 
course of their first oscillation. These PBH's may then decay to leave topologically stable Planck-mass relics, though this possibility has 
not been thoroughly investigated \cite{Avgoustidis:2005vm, Matsuda:2004bg}.

By contrast Matsuda has claimed that a large fraction $f \sim 1$ of all necklace loops eventually collapse to form PBH's via a separate necklace-specific process. 
This claim is based on the assumption that mass-energy may only be lost, via gravitational wave emission, from the four-dimensional string segments, 
leaving the bead mass unchanged. Thus when the necklace loop reaches its minimum radius it will undergo collapse if the mass of the beads is large 
enough to produce a Schwarzschild radius greater than the string width.\footnote{We assume here that the minimum radius of the loop is 
determined by the effective width of the string.} He therefore proposed that small loops produced at very early times may form stable relics which also 
act as dark matter candidates. Conversely he argued that loops created at later times would be larger and hence likely to contain enough mass in their beads to 
cause them to collapse into PBH's \cite{Matsuda:2006ju}.

However this original schematic analysis involved a number of simplifying assumptions, such as the existence of a time-independent lifting potential and hence a 
constant bead mass for necklace loops formed at different epochs. The initial inter-bead spacing was also assumed to scale like the entropy distance at the 
time of network formation\footnote{$d_S(t_s) \sim t_M^{\frac{1}{2}}t_s^{\frac{1}{2}}$ \cite{book}, where $t_s$ is the formation time of the string 
network, $t_M$ is the time of monopole formation such that $t_M \leq t_s$.} and the dynamical evolution of the inter-bead distance was modeled by the 
standard string-monopole network evolution equations, originally proposed by \cite{Berezinsky:1997td}. 

In the following analysis we attempt to construct a more concrete model of necklace formation based on 
ideas from type IIB string theory. We calculate the explicit form of the lifting potential for a loop of string with extra-dimensional windings in the 
Klebanov-Strassler geometry \cite{Klebanov:2000hb, Herzog:2001xk}. Using realistic models of winding and loop formation 
we see that potential itself evolves dynamically resulting in a time dependent bead mass. 
This shows that the first of these assumptions must be modified, at least for certain backgrounds. 

Additionally the decay signature of necklace loops in our model is in many ways the opposite of what Matsuda predicted. 
We find that PBH formation is favoured at early times with potential dark matter relics forming later. 
This is indeed an unexpected result \cite{Matsuda:2005ez}, though one which appears to follow naturally from 
the consideration of monopole/bead formation as a \emph{dynamical} process rather than as the result of a separate phase transition prior to string formation. 
We argue that this is the correct approach to take when considering monopoles which form from extra-dimensional 
windings and a comparison of our results with field-theoretic string-monopole networks is given in section 5.6. 

In addition to a renewed interest in the role of $F$/$D$-string networks in cosmology there are still many in the physics community devoted to the study of 
field theoretic cosmic strings \cite{Hindmarsh:1994re, Copeland:2005cy, Polchinski:2007qc}. 
Whilst a stringy origin of the CMB perturbations has been ruled out by observation, 
the best $\Lambda$ CDM model fit to the data suggests that these strings may contribute at the level of $\sim 10\%$, \cite{Nevis:2007} (for a review see also 
 \cite{Sakellariadou:2009ev} )
making them extremely important objects to study. 
Unfortunately a best fit for $(p,q)$ strings, necklaces or cycloops has not yet been investigated, at least at the field theory level \cite{Rajantie:2007hp}. 

However recent discoveries of dualities between gauge field strings and $F/D$-strings have raised the possibility of unifying these two approaches.
Indeed, if string theory really is a theory of everything (TOE), and if we accept Quantum Field Theory (QFT) to be a valid low-energy approximation, 
we may also hope to find string theory analogues of all field theory phenomena. 
If therefore we expect topological defects, including strings, to arise generically in symmetry breaking processes, 
we must investigate the relationship between fundamental strings and field theory strings in much more detail.    

The aim of this paper is to consider a simple model of string necklaces in a well understood supergravity background using type IIB string theory, 
thereby extending the initial phenomenological approach begun in \cite{Matsuda:2005ez}.
Following the considerations above therefore we also consider the possible relation of these objects to field theoretic strings, restricting ourselves to 
generic considerations. Since the background yields an explicit form for the uplifting potential in terms of the number of extra-dimensional windings we may 
compute the bead mass precisely if this number is known for a string loop formed at any epoch. Following \cite{Avgoustidis:2005vm} we assume that the 
motion of a string in the compact space, prior to the chopping off of a loop, is random. 
This therefore allows us to estimate both the time-dependent bead mass and the average inter-bead distance. 
We find that the results depend the definition of the parameter $\omega_l$, which gives the fraction of the total string 
length contained in the extra-dimensional windings. Two definitions are suggested: 
Identifying the inter-bead distance in the string picture with the correlation length in the field theory picture, 
we see that the first definition leads naturally to a scaling solution - similar to that for field theoretic strings but with a 
correlation length $L(t) \sim \gamma t$ much smaller than the horizon. The second leads to sub-scaling solution $L(t) \sim t^{\frac{3}{4}}$.

More importantly we are able to compute the PBH mass spectrum produced by collapsing necklaces. 
Since the mass of an individual necklace depends upon the structure of the internal manifold as well as the string tension, 
the resulting PBH spectrum yields information about the size of the of the extra dimensions and the warp factor. 
This in turn influences the background cosmic ray flux via the Hawking radiation of PBH's expiring at the present epoch. We are thus able to provide 
observational bounds on string theory parameters using measurements of the extra-galactic gamma-ray flux at 100MeV from the EGRET experiment \cite{Hartman:1999fc}. 

An interesting result is that both definitions of $\omega_l$ give the same qualitative behaviour with PBH formation occurring only over a 
limited time period in the early universe. However it must also be noted that the upper and lower limits of this window 
and hence the resulting bounds on the model parameters do vary significantly in each case.

The layout of the paper is then as follows: In Section 2 we introduce the Klebanov-Strassler geometry which is the supergravity background for the
string embedding. In Section 3 we construct the world-volume action for string loops before analysing their stability in Section 4. Section 5 deals with the formation
of these loops and their cosmological impact, focusing on the predictions for PBH abundance and Section 6 contains a brief discussion of our main results and 
suggestions for future work.

We conclude this introduction with a note regarding terminology.  
The "necklaces" which are the subject of the present paper and of Matsuda's original work should not be confused with "necklaces" 
formed via other string-monopole interactions. For example, Leblond and Wyman \cite{Leblond:2007tf} have shown that $D0$-branes (monopoles) 
may be formed at junctions between strings in a $(p,q)$-string network. The resulting "necklaces" are of no relation to 
the ones considered here. Related work can be found in \cite{Rocha:2008de, Leblond:2009fq, Bevis:2009az, Abbott:2009rr}.

%%%%%%%%%%%%%%%%%%%%%%%%%%%%%%%%%%%%%%%%%%%%%%%%%%%%%%%%%%%%%%%%%%%%%%%%%
%%%%%%%%%%%%%%%%%%%%%%%
\section{Klebanov-Strassler geometry}\label{conifold}
The background we wish to consider is that of the Klebanov-Strassler (KS) throat \cite{Klebanov:2000hb, Herzog:2001xk} since it is one of the better understood 
backgrounds of the type IIB theory. Recall that conical singularities are the most generic kind of singularities arising within compactifications 
of IIB string theory on manifolds of $SU(3) \times SU(3)$ structure \cite{Grana:2005jc}. Since explicitly realistic compactifications are difficult to construct, 
we will take a more phenomenological approach by considering (non-compact) conical backgrounds such as the conifold $T^{1,1}$, 
that can be glued to a (conformal) Calabi-Yau manifold. Provided we work in a region far from this gluing, we can (locally) work with the 
conifold geometry without worrying too much about the precise details of the compactification mechanism. With this in mind we can regularise the singular 
conifold by allowing the $S^2$ to shrink to zero size. This is just the deformation of the conifold, which is topologically 
equivalent to the cotangent bundle over the three-sphere $T^{*}S^3$ where the $S^3$ has some minimal size.

The solution of Klebanov-Strassler requires the introduction of both $D3$-branes and fractional $D3$-branes in this background \cite{Klebanov:2000hb}, 
where the fluxes back-react on the geometry to create a warped throat. 
The $RR$-three form flux is threaded through the finite size three-sphere, whilst the $NS$-$NS$ flux wraps the dual (B) cycle as follows,
%(1.1)
\begin{equation}
\frac{1}{ 4\pi^2 \alpha'} \int_{S^3} F^{(3)} = M, \hspace{1cm} \frac{1}
{4\pi^2 \alpha'} \int_{B} H^{(3)} = -K.
\end{equation}
The fluxes generate a non-trivial warp factor for the full ten-dimensional solution which effectively measures the size of the blow-up contribution along the $S^2$ via
%(1.2)
\begin{eqnarray}
h(\tau) &=& 2^{2/3}(g_s M \alpha')^2 \epsilon^{-8/3}I(\tau) \nonumber \\
I(\tau) &=& \int_{\tau}^{\infty} \frac{x \coth(x)-1}{\sinh^2(x)} \left(\sinh(2x)-2x \right) dx.
\end{eqnarray}
For the solution we are considering, we must send $\tau \to 0$, since $\tau$ controls the radius of the $S^2$, and therefore
the warping approaches a constant value $a_0 =\rm{Lim}_{\tau \to 0} h(\tau)$, and the ten-dimensional metric is well
approximated by the following;
%(1.3)
\begin{equation}\label{eq:metric}
ds^2 = a_0^2 \eta_{\mu \nu} dx^{\mu} dx^{\nu} + R^2 (d\psi^2 + \sin^2 \psi (d\theta^2 + \sin^2 \theta d\phi^2))
\end{equation}
with $B^{(2)} \sim C^{(0)} \sim 0$ at the tip, and 
%(1.4)
\begin{equation}
C^{(2)} = \frac{1}{g_s} M \alpha' F(\psi) \omega_2
\end{equation}
where $\omega_2$ is the volume form along the two-cycle parameterised by $\theta, \phi$, and $F(\psi) \sim \psi-\sin\psi \cos \psi$.
In this definition we are explicitly assuming the background gauge choice \cite{Firouzjahi:2006vp}
%(1.5)
\begin{eqnarray}
\theta &=& \theta_1 = - \theta_2 \nonumber \\
\phi &=& \phi_1 = -\phi_2
\end{eqnarray}
and $\psi$ is now the azimuthal angle on the $S^3$ with $\psi \in (0, 2\pi]$ as its fundamental domain. 
Note also that the $x^{\mu}$ are dimensionful coordinates whilst we define the radius of the three-sphere via,
%()
\begin{equation}
R^2 = bM g_s  \alpha'
\end{equation}
where $b$ is a numerical factor and $g_s$ 
is the string coupling, keeping the angular variables dimensionless. It is important to note that the dilaton, and therefore the string coupling, is constant
in this background.

In what follows we will assume that (\ref{eq:metric}) is representative of a large class of warped geometries, 
without necessarily having explicit realisation in a fully UV complete construction. In particular this means that we should treat the warping parameter $a_0$
as a constant satisfying $0 < a_0 < 1$. For the (non-compact) KS solution we find that $a_0$ is related to the deformation parameter of the conifold $\epsilon$
via the expression
\begin{equation}
a_0^2 \sim \frac{\tilde{\epsilon}^{4/3}}{R^2}
\end{equation}
where $\tilde{\epsilon}$ is re-scaled to have the correct dimensions and is related to the deformation parameter $\epsilon$ of the 
warped conifold. The basic point is that the warping varies like $1/g_sM$ where we are assuming the supergravity
(SUGRA) limit $g_sM >> 1$ so that, for fixed $R$, constraints on the warping can be interpreted as directly constraining the background geometry.
We should also point out that there is nothing special about the deformed conifold solution. One could equally well
use the resolved conifold where the $S^2$ is blown up instead \cite{Klebanov:2007us, Klebanov:2007cx}. 
Indeed the resulting cosmic-string tension scales directly with the resolution parameter and is therefore highly constrained by observations.
The resulting solution is then very similar to the model considered in \cite{BlancoPillado:2005dx}.

Our choice for the background is also inspired in part by the AdS/CFT duality, since the KS geometry is known to be dual to an $\mathcal{N}=1$
confining gauge theory. When viewed from this perspective, the strings are effectively the confining strings of the gauge theory with non-trivial wrapping. 
Whilst this is interesting in its own right, our motivation will be to model cosmic-necklaces rather than gauge theory necklaces.

%Section2%%%%%%%%%%%%%%%%%%%%%%%%%%%%%%%%%%%%%%%%%%%%%%%%%%%%%%%%%%%%%%%%
%%%%%%%%%%%%%%%%%%%%%%%
\section{String loops with non-trivial windings in internal space}\label{loops}

The action for both fundamental strings ($F$-strings) and $D1$-branes ($D$-strings) - or in more general $(p,q)$ notation $(1,0)$ and $(0,1)$ strings respectively - in the warped deformed conifold is simply the Nambu-Goto action with additional world-sheet flux, plus a possible Chern-Simons term (for the $D$-string),
%(2.1)
\begin{equation}
S = -T_1 \int d\sigma dt \sqrt{-X} + S_{CS}  
\end{equation}
where $T_1$ is the string tension\footnote{Note that in the warped throat geometry we do not expect the tensions of the F and D-strings to be exactly equal, $T_{(1,0)} \neq T_{(0,1)}$. The formula for the tension of a general $(p,q)$ string, in the large $p$/large $q$ limit, in the warped throat background, is $T_{(p,q)} \approx \frac{1}{2\pi \alpha'}\sqrt{\left(\frac{q}{g_s}\right)^2 + sin^2\left(\frac{p\pi}{M}\right)}$, but it remains an open problem to calculate the tension for small $p$/small $q$. We may however assume that the order of magnitude of \emph{both} the F and D-string tensions are set by the fundamental string scale, i.e. $T_{(1,0)} \sim T_{(0,1)} \sim \alpha'^{-1}$. Here we will use the term $T_1$ to refer to either tension.}
%Here however we use the term $T_1$ to denote either tension. The two 
%cases with be dealt with separately in Section \ref{cosmology}.} 
and $X = \rm{det}{X_{ab}}$,
%(2.2)
\begin{equation}
X_{ab} = \gamma_{ab} + \lambda F_{ab}
\end{equation}
where $a,b \in \left\{0, \sigma \right\}$, $\gamma_{ab}$ is the usual 
induced metric on the world-sheet,
%(2.3)
\begin{equation}\label{eq:inducedmetric}
\gamma_{ab} = G_{MN}(X(\sigma,t)) \partial_a X^M(\sigma,t) \partial_b X^N(\sigma,t)
\end{equation}
and $\lambda = 2 \pi l_s^2$.
The simple form of the Lagrangian density $\sqrt{-X}$ arises because we are neglecting the coupling of the world-sheet to the $NS$-$NS$ two-form field, 
which is vanishing in our background (at least in the limit we are considering). 
The flux tensor is anti-symmetric such that $\lambda F_{00} = \lambda F_{\sigma \sigma} = 0$  and 
$\lambda F_{0 \sigma} = -\lambda F_{\sigma 0}$, $\lambda F_{0 \sigma} \geq 0$. Note that we are absorbing the definition of the string coupling into 
the field strength tensor, since this allows us to identify the String frame with the Einstein frame.
The Chern-Simons coupling is given by the integral of the pull-back of the $RR$ two-form over the world-sheet, for the case of D-strings,
%(2.4)
\begin{equation}
S_{CS} = T_1 \int d\sigma dt M \alpha' (\psi - \sin\psi \cos\psi) d\theta \wedge d\phi   
\end{equation}
However, for the sake of simplicity we choose to ignore this correction in the following analysis and instead concentrate solely on the Nambu-Goto 
component of the action.

We wish to take the following general ansatz for the string embedding,
\begin{equation}\label{eq:embeddingansatz}
X^M = (t, r(t)\sin \sigma, r(t)\cos\sigma, z_0, \tau \rightarrow 0, 0,0 ,\psi(\sigma,t), \theta(\sigma,t), \phi(\sigma,t))
\end{equation}
where $0 < \psi \leq 2\pi$, $0 < \theta \leq \pi$ and $0 < \phi \leq \pi$. This describes a circular string in the Minkowski directions which is wrapped
\emph{smoothly} over the $S^3$ in the internal space. Note that in this model the strings sit at the tip of the warped throat. We also make this choice for the sake of simplicity. However there
is no a priori reason why we should make this assumption. The more general case would follow the lines of \cite{Avgoustidis:2007ju}.
Using the metric (\ref{eq:metric}) and the embedding ansatz (\ref{eq:embeddingansatz}) the action then becomes
%(2.6)
\begin{equation}\label{eq:stringaction}
S = -T_1 a_0^2\int d\sigma dt  \sqrt{(1-\dot{r}^2)(r^2+a_0^{-2} R^2 s^{\prime 2})-a_0^{-2} r^2 R^2 \dot{s}^2-a_0^{-4} 
\lambda^2 F^2_{0 \sigma} -a_0^{-4} R^4 \dot{s}^2 s'^2 + a_0^{-4} R^4 (\dot{s}s')^2} 
\end{equation}
where we have introduced the slightly abusive notation,  
%(2.7) and (2.8)
\begin{eqnarray}
\dot{s}^2 &=& \dot{\psi}^2 + \sin^2\psi(\dot{\theta}^2 + \sin^2\theta \dot{\phi}^2) \nonumber \\
s'^2 &=& \psi'^2 + \sin^2\psi(\theta'^2 + \sin^2\theta \phi'^2) \\ 
(\dot{s}s') &=& \dot{\psi} \psi' + \sin^2\psi (\dot{\theta} \theta' + \sin^2\theta \dot{\phi} \phi') \nonumber
\end{eqnarray}
where a dot or dash indicates differentiation with respect to $t$ or $\sigma$ respectively and we have chosen the gauge so as to identify 
the world-sheet time coordinate with the proper time in the Lorentz frame of the string loop, $t$.
We have also included a non-zero 'electric' gauge field contribution for 
generality. For the $D$-string case this corresponds to the dissolving of $F$-string 
charge on the world-sheet and the strings are essentially superconducting.

%Explicitly in the case of the $D$-string, the Chern-Simons term becomes
%(2.10)
%\begin{equation}
%S_{CS} = T_1 \int d\sigma dt M \alpha' (\psi - \sin\psi \cos\psi) \sin\theta \epsilon^{ab}\partial_a \theta \partial_{b} \phi
%\end{equation}
%where $a,b$ run over the world-sheet coordinates.
There are two constants of motion for this configuration, the total energy of the string $H$ and the angular momentum $l$, due to the motion of the
string in the internal dimensions\footnote{Note that although the string "rotates" around the $S^3$, no centripetal force is acting upon it. The internal (compact) 
dimensions are parameterised in terms of the angular variables $\psi$, $\theta$ and $\phi$ and so the motion through the $S^3$ is measured in rad x $[t]^{-1}$, 
the units of angular velocity.}${}^{,}$\footnote{We have not here considered the more general case of a loop with non-trivial windings in the 
internal manifold which also rotates in Minkowski space. Such a scenario would require the action of a genuine centripetal force provided by the string effective tension.}.
We parameterise these conserved charges as follows \cite{BlancoPillado:2005dx}
%(2.11)
\begin{equation}
H = \frac{\partial L}{\partial \dot{q}^I} \dot{q}^I - L, \hspace{0.6cm} l = \frac{\partial L}{\partial \dot{q}^I} q'^I
\end{equation}
where $L$ is the Lagrangian and $q^I$ are the canonical coordinates which are themselves functions of the world-volume coordinates through
$q^I = (r(t), \sigma, \psi(\sigma,t), \theta(\sigma,t), \phi(\sigma,t))$. Using (\ref{eq:stringaction}) and neglecting the Chern-Simons term, we find the 
following expressions for these conserved charges
%(2.13)
\begin{eqnarray}
H &=& T_1 \int d\sigma \frac{a_0^2 (r^2+a_0^{-2}R^2 s'^2)}{ \sqrt{(1-\dot{r}^2)(r^2+a_0^{-2} R^2 s^{\prime 2})-a_0^{-2} r^2 R^2 \dot{s}^2-a_0^{-4} \lambda^2 F^2_{0 \sigma} - a_0^{-4} R^4 \dot{s}^2 s'^2 + a_0^{-4} R^4 (\dot{s}s')^2}} \nonumber \\
l &=& T_1 \int d\sigma \frac{R^2 r^2(\dot{s}s')}{\sqrt{(1-\dot{r}^2)(r^2+a_0^{-2} R^2 s^{\prime 2})-a_0^{-2} r^2 R^2 \dot{s}^2-a_0^{-4} \lambda^2 F^2_{0 \sigma} - a_0^{-4} R^4 \dot{s}^2 s'^2 + a_0^{-4} R^4 (\dot{s}s')^2}} \nonumber
\end{eqnarray}
However it is more useful to re-write the Hamiltonian in canonical form using the momenta;
%(2.15)
\begin{eqnarray}
P^2 &=& \left(\frac{\partial L}{\partial \dot{r}}\right)^2, \hspace{0.5cm} \Pi^2 = \left(\frac{\partial L}{\partial \dot{A}}\right)^2 \nonumber\\
L_{\psi}^2 &=& \left(\frac{\partial L}{\partial \dot{\psi}}\right)^2, \hspace{0.5cm} L_{\theta}^2 = \left(\frac{\partial L}{\partial \dot{\theta}}\right)^2, \hspace{0.5cm} 
L_{\phi}^2 = \left(\frac{\partial L}{\partial \dot{\phi}}\right)^2 
\end{eqnarray}
This will simplify the form of our solutions for stable string windings with non-zero world-sheet flux $\lambda F_{0 \sigma} \neq 0$. 
This is especially true in the static $(l = 0)$ case which we will consider in Section \ref{stability}.

After a long but straight forward calculation we find that the canonical form of the Hamiltonian reduces to,
%(2.16)
\begin{equation}\label{eq:canonicalhamiltonian}
H =  \int d\sigma \sqrt{r^2 + a_0^{-2}R^2 s'^2}\sqrt{T_1^2 a_0^4 + \frac{P^2}{(r^2 + a_0^{-2}R^2 s'^2)} + \frac{\Pi^2}{a_0^{-4} \lambda^2} + \tilde{L_\psi^2} + \tilde{L_\theta^2} + \tilde{L_\phi^2}}
\end{equation}
where we have written
%(2.17)
\begin{eqnarray}\label{eq:psimomentum}
\tilde{L_{\psi}}^2 &=& \frac{a_0^4 \dot{\psi} L_\psi^2}{T_1^2 a_0^2 R^2 {r^2\dot{\psi} + \sin^2\psi[\theta'(\dot{\psi}\theta' - \dot{\theta}\psi') + \sin^2\theta \phi'(\dot{\psi}\phi' - \dot{\phi}\psi')]}} \nonumber\\
\tilde{L_{\theta}}^2 &=& \frac{a_0^4 \dot{\theta} L_\theta^2}{T_1^2 a_0^2 R^2\sin^2\psi{r^2\dot{\theta} + [\psi'(\dot{\theta}\psi' - \dot{\psi}\theta') + \sin^2\psi \sin^2\theta \phi'(\dot{\theta}\phi' - \dot{\phi}\theta')]}} \\
\tilde{L_{\phi}}^2 &=& \frac{a_0^4 \dot{\phi} L_\phi^2}{T_1^2 a_0^2 R^2 \sin^2\psi \sin^2\theta {r^2\dot{\phi} + [\psi'(\dot{\phi}\psi' - \dot{\psi}\phi') + \sin^2\psi (\dot{\phi}\theta' - \dot{\theta}\phi')]}} \nonumber
\end{eqnarray}
In the next section we will specify the ansatz (\ref{eq:embeddingansatz}) completely by identifying the functions $\psi(\sigma,t)$, $\theta(\sigma,t)$ and $\phi(\sigma,t)$. 
We will then use equation (\ref{eq:canonicalhamiltonian}) to determine the stability conditions in the static case by minimising the total energy of the string. 
Finally we see how this leads naturally to a concrete model of necklace loops in the warped throat scenario when these solutions are perturbed.
%Section3%%%%%%%%%%%%%%%%%%%%%%%%%%%%%%%%%%%%%%%%%%%%%%%%%%%%%%%%%%%%%%%%
%%%%%%%%%%%%%%%%%%%%%%
\section{Stability analysis for string loops in the static $(l=0)$ case and necklace formation}\label{stability}

In the static case we assume that there is no motion in the compact dimensions and that the loop is neither expanding nor contracting in Minkowski space. 
Setting $\dot{\psi} = \dot{\theta} = \dot{\phi} = 0$ and $\dot{r} =0$ (or equivalently $L_\psi^2 = L_\theta^2 = L_\phi^2 = 0$ and $P^2 = 0$) gives the static 
potential $V$ (below) and $l=0$ as expected.
%(3.1)
\begin{equation}\label{potential}
V = \int d\sigma a_0^2 \sqrt{T_1^2 + \frac{\Pi^2}{\lambda^2}} \sqrt{r^2+ a_0^{-2}R^2 s'^2}
\end{equation}
Note that even static strings of this kind may lose mass-energy over time due to the emission of gravitational radiation, causing them to shrink. 
However we refer here to the stability of extra dimensional-windings in the static gauge over an epoch in which the loop size $r$ and energy $V$ are roughly constant. 
The shrinking of necklace loops over cosmological time scales and its implications are dealt with explicitly in Section \ref{cosmology}.

We must now complete the ansatz (\ref{eq:embeddingansatz}). Choosing $\psi(\sigma,t)$, $\theta(\sigma,t)$, and $\phi(\sigma,t)$ to be such that the 
angular winding is linear in $\sigma$ \footnote{This configuration corresponds to a situation in which the end 
point of the string is equally likely to move in each of the angular directions prior to the moment of loop formation. As we shall see in 
Section \ref{cosmology}, when we consider the random walk regime, this assumption 
is well motivated from a physical point of view. We also note for future 
reference that in canonical coordinates, windings of this form do not wrap 
geodesics in the $S^3$. This has important implications for the $\sigma$-dependence of both $H$ and $l$. }, we see
%(3.2),(3.3) & (3.4)
\begin{eqnarray}\label{eq:angularwindings}
\psi(\sigma,t) &=& 2n_{\psi}\sigma + \psi(t),\nonumber \\
\theta(\sigma,t) &=& n_{\theta}\sigma + \theta(t),\\
\phi(\sigma,t) &=& n_{\phi}\sigma + \phi(t) \nonumber
\end{eqnarray}
where $n_{\psi}$, $n_{\theta}$, $n_{\phi} \in N$ and $0 < \sigma \leq 2\pi$ then gives
%(3.5)
\begin{equation}\label{eq:effectivepotential}
V = \int d\sigma a_0^2 T_1 \sqrt{1 + \frac{\Pi^2}{T_1^2 \lambda^2}} \sqrt{r^2+ a_0^{-2}R^2(4n_\psi^2 + \sin^2\psi(n_\theta^2 + \sin^2\theta n_\phi^2))}
\end{equation}
where $\psi$ and $\theta$ are themselves functions of $\sigma$ and $t$ 
according to (\ref{eq:angularwindings}).

We see immediately that $V = V(\psi,\theta)$ indicating that the $\phi$-direction is flat, whereas the $\psi$ and $\theta$ directions are "lifted" 
by the presence of a potential energy density $\mathcal{V}$ which is the integrand in (\ref{eq:effectivepotential}).

It is now possible to make a connection between wound strings at the tip of the conifold throat, and the cosmic necklaces predicted generically by Matsuda 
\cite{Matsuda:2004bg}. 
First we must determine the conditions under which the wrappings described above are stable, which is done by minimising the total energy $V$. 
Perturbations of this configuration are then seen to give rise to beads, 
whose mass can be estimated from the functional form of the potential.

However rather than simply minimising the potential, it is useful (and easier) at this point to develop a physical intuition for the string configuration.
Treating $\psi$ and $\theta$ as independent variables, we may sketch the lifting potential (strictly speaking the lifting potential energy density, 
but from here on these two terms will be used interchangeably) which the string "sees" throughout the whole $S^3$. 
This is done by plotting the integrand $\mathcal{V}$ in (\ref{eq:effectivepotential}). 
The result for the principle range ($0 < \psi \leq 2\pi$, $0 < \theta \leq \pi$) is shown in Figure 1 for the purposes of illustration.
%Figure 1
\begin{figure}[htp]
 \begin{center}
  \includegraphics[width=0.6\textwidth]{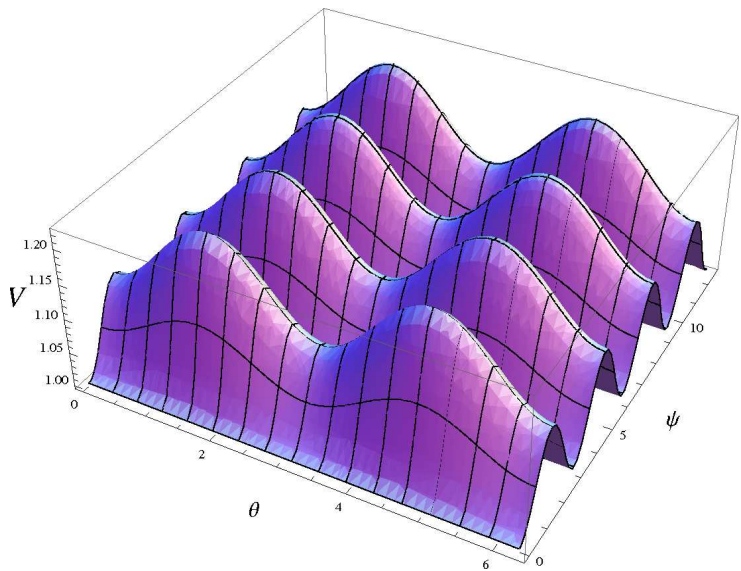}
  \caption{Plot of the static potential $\mathcal{V}$ setting the tension pre-factor term to unity and also $r^2 = 1$ and $a_0^{-2}R^2 = 1$.}
 \end{center}
\end{figure}
The critical points and associated field masses may be found by diagonalising the corresponding Hessian matrix. 
We verify that $\mathcal{V}$ has equal local maxima at positions $(\psi =\frac{(2n+1)\pi}{2}, \theta = \frac{(2m+1)\pi}{2})$, 
saddle points of equal magnitude at $(\psi = \frac{(2n+1)\pi}{2}, \theta = m\pi)$ and flat directions 
which are also local minima (i.e. "troughs" not "ridges" or points of inflection) given by $(\psi = n\pi, \theta)$, where $m,n \in Z$. 
This gives rise to two local maxima, two saddle points and two flat directions in the principle range. The associated field masses are given below.
%(3.6)
\begin{itemize}
\item $\psi = \frac{(2n+1)\pi}{2}$, $\theta = \frac{(2m+1)\pi}{2}$ (local maxima)
	\begin{eqnarray}
	m_{\theta}^2  = -\frac{T_1 R^2 n_{\phi}^2}{\sqrt{r^2 + a_0^{-2} R^2 (4
n_\psi^2 + n_\theta^2 + n_\phi^2)}},
	m_{\psi}^2 = -\frac{T_1 R^2 (n_{\theta}^2 + n_\phi^2)}{\sqrt{r^2 + a_0
^{-2} R^2 (4n_\psi^2 + n_\theta^2 + n_\phi^2)}} 
	\end{eqnarray}
%\end{itemize}
%(3.7)
%\begin{itemize}
\item $\psi = \frac{(2n+1)\pi}{2}$, $\theta = m\pi$ (saddle points)
\begin{eqnarray}
m_{\theta}^2  = \frac{T_1 R^2 n_{\phi}^2}{\sqrt{r^2 + a_0^{-2} R^2 (4n_\psi^2 + n_\theta^2)}},
m_{\psi}^2  = -\frac{T_1 R^2 n_{\theta}^2}{\sqrt{r^2 + a_0^{-2} R^2 (4
n_\psi^2 + n_\theta^2)}} 
	\end{eqnarray}
%\end{itemize}
%(3.8)
%\begin{itemize}
\item $\psi = n\pi$, $\theta$ (flat directions)
\begin{eqnarray}
m_{\theta}^2 = 0 ,
m_{\psi}^2  = \frac{T_1 R^2 (n_{\theta}^2 + sin^2\theta n_\phi^2)}{\sqrt{r^2 + 4a_0^{-2} R^2 n_\psi^2}} 
\end{eqnarray}
\end{itemize}
It is now intuitively clear that minimal energy configurations correspond to strings wrapping flat directions in the two-dimensional sub-manifold described by 
$\psi$ and $\theta$. Physically this corresponds to strings wrapping some point in the $S^3$ which is uniquely determined by the condition $\psi = n\pi$ for some $n\in Z$. 
This may be seen from the metric (\ref{eq:metric}) and is the reason why $n_\theta$ and $n_\phi$ do not contribute to the total energy. 
Even though, technically, $n_\theta$,$n_\phi \geq 0$ windings around points have zero length and are not physically meaningful.
This result is precisely what we should expect, since the minimum energy 
configuration ought to correspond to a situation in which the string has effectively zero length involved in extra-dimensional windings. 

Indeed, we may verify this by substituting $\psi = n\pi$ and setting $\Pi^2 = 0$ in (\ref{eq:effectivepotential}) that the total energy of a string wrapping
flat directions in the potential $\mathcal{V}$ is given by
%(3.9)
\begin{equation}
V = 2\pi a_0^2 r T_1
\end{equation} 
which is simply the rest mass of a string loop with radius $r$ in warped Minkowski space.

This is certainly a very complicated way of verifying an intuitively obvious result but the above analysis will prove useful 
when we consider perturbations which result in different sections of a single string lying along equivalent local minima. 
That is, when the string interpolates between degenerate minima (flat directions) in the $(\psi, \theta)$ sub-manifold resulting in the formation of beads.

Note that the flat directions in Figure 1 effectively define degenerate vacuum states for the string. 
For this reason we propose identifying the inter-bead distance in the string picture with the correlation length of field-theoretic strings. 
This forms the basis for identifying the field-theoretic parameter $\gamma$ (which defines the correlation length as a fraction of the horizon distance) 
with the parameters which define the KS geometry in the following section. 
Note also that, although we refer to the bead-forming states as perturbations from the minimum energy (zero winding) configuration, 
this is true only in an energetic sense. Such configurations differ locally from the minimum energy (zero winding) configuration only in the vicinity of a bead. 

Indeed it is unclear whether it is possible to create a necklace from a standard $F$ or $D$-string loop via physical perturbations of a 
section of string \emph{after} the formation of the loop. 
This is the same as asking whether it is meaningful for an open string section (which may itself form part of a string loop or part of a 
string section connected to the network) to contain fractional windings which result in the formation of beads.

It is subtle point but in his generic argument Matsuda \cite{Matsuda:2004bg} predicted that, 
due to the presence of the lifting potential, integer windings would \emph{not} be necessary for stability. In principle this should allow 
for the formation of beads in open string sections due to local fractional windings. Though there is no reason why this could not happen generically, the 
possibility is not explicitly realised in our model. As discussed in the following section, we assume that any small-scale structure due to local partial windings 
will quickly disappear due to the annihilation of beads/anti-beads.

In our model the existence of the potential barrier which stabilises the windings after the loop chops off from the network 
(and which allows for the formation of beads) depends upon the presence of non-zero integer windings 
\emph{at the moment of loop formation}. What is more, this ought to be true generically in this class of models since the form of the lifting 
potential in the extra dimensions ought always to be determined by the parameters that characterise the internal windings.

Physically what happens is the following; before the loop chops off from the network the string is free to move and create windings in the compact dimensions. 
If the internal manifold were not simply connected these would become topologically trapped resulting in the formation of cycloops \cite{Avgoustidis:2005vm}. 
However as the $S^3$ is simply connected, these windings cannot be topologically stabilised. 
Instead they are stabilised by the presence of the lifting potential $\mathcal{V}$, which is itself a function of the number of windings.

We may imagine that, at the moment of loop formation the string ansatz is described accurately by (\ref{eq:angularwindings}) which is sketched in Figure 2a. 
However as soon as the string chops off from the network to form a loop, the total energy ($V$) is no longer minimised for such a smoothly varying configuration. 
Although windings may continue to vary smoothly in the $\phi$-direction they will then adopt a step-like configuration in the $(\psi, \theta)$ sub-manifold 
which is illustrated in Figure 2b.

Note that this is an approximation. Technically if the string configuration (i.e. the ansatz) changes, then the form of the lifting potential also changes. 
This results in a complicated iterative process with complicated string dynamics before the eventual formation of a steady state.

However if we consider $\mathcal{V}$ to be approximately constant, we see that the integral $V = \int d\sigma \mathcal{V}$ is minimised precisely 
for the step-like configuration shown in Figure 2b in which the string interpolates 
between degenerate minima by crossing the potential barrier perpendicularly at its lowest point i.e. at a saddle point $\psi = \frac{(2n+1)\pi}{2}$, $\theta = m\pi$. 
This in turn may be thought of energetically as a series of small perturbations away from the genuine minimum energy configuration of zero-windings (described above) 
which is equivalent to a string wrapping a single flat direction in the $(\psi,\theta)$ sub-manifold. 
Looking at it from this perspective helps us to justify our original assumption that $\mathcal{V}$ remains approximately constant, 
so long as we remember to set $n_\psi = 0$ in (\ref{eq:effectivepotential}).
%Figure 2
\begin{figure}[htp]
 \begin{center}
  \includegraphics[width=1\textwidth]{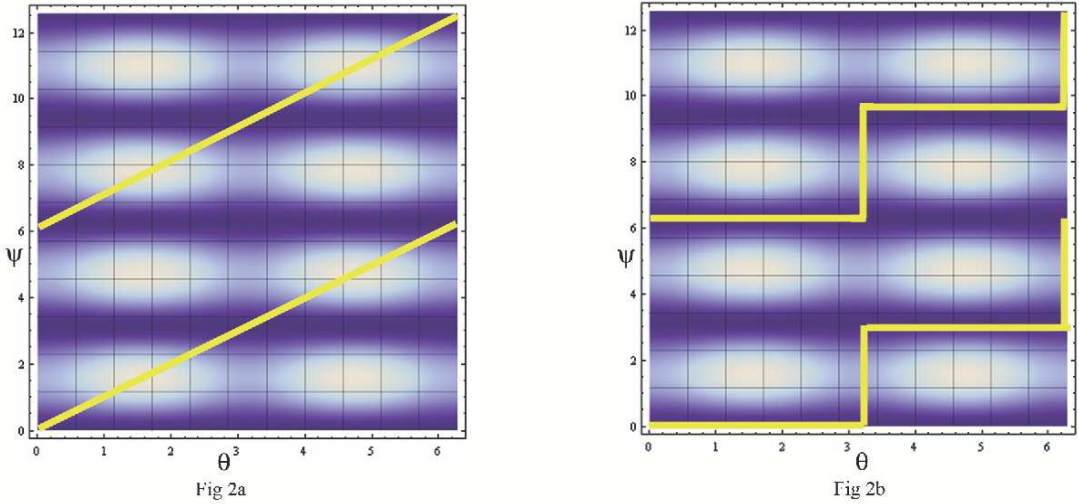}
  \caption{Figure 2a shows strings winding smoothly across the effective potential $\mathcal{V}$ on the $(\theta,\phi)$ sub-manifold. 
Figure 2b shows step-like configurations which minimises the total string energy within $\mathcal{V}$.}
 \end{center}
\end{figure}
This leads to an apparent contradiction. We may assume, at the moment of loop formation, that $n \sim n_\psi \sim n_\theta \sim n_\phi > 0$. 
After this point windings in the $\phi$-direction will still be able to vary smoothly, though they will not be stable and may in principle contract 
until they are point-like. This possibility is discussed in detail in section 5.6.

However, so long as $n_\phi > 0$, windings in the ($\psi$, $\theta$) sub-manifold will relax into the step-like configuration shown in Figure 2b 
resulting in the formation of approximately $2n_\psi$ beads. This forces us to view the configuration as a series of small perturbations 
(in fact a series of $2n_\psi$ perturbations) away from a configuration described by $\psi = n\pi$ and $n_\psi = 0$. We must therefore continue to 
regard $n_\psi > 0$ (the number of windings in the $\psi$-direction at the moment of loop formation) 
as physically meaningful when estimating the number of beads, but we must regard it as approximately zero when estimating the bead mass. 
This is a little strange but it is quite consistent with the physical picture we have been sketching.

To obtain our estimate for the bead mass ($M_b$) therefore we first Taylor expand (\ref{eq:effectivepotential}) with $n_\psi$ set equal to zero. 
Assuming $a_0r >> nR$ at the moment of loop formation, which corresponds to a large loop in the Minkowski directions,  for all times we 
obtain the approximate expression,
%(3.10)
\begin{equation}
V \sim  \sqrt{1 + \frac{\Pi^2}{T_1^2 \lambda^2}} \left(2\pi a_0^2 r T_1 + T_1 \int_{0}^{2\pi} d\sigma \frac{R^2 \sin^2\psi(n_\theta^2 + \sin^2\theta n_\phi^2)}{ r}\right)
\end{equation}
We then move from the smooth windings picture shown in Figure 2a to the step-like configuration in Figure 2b by setting $\psi=m\pi$ 
globally, except in the vicinity of a bead where $\theta=m\pi$, $d\psi = d\sigma$ and $m\pi \leq \psi \leq (m+1)\pi$. Our expression for the total energy $V$ is then, 
%(3.11)
\begin{eqnarray}\label{eq:totalpotential}
V \sim  \sqrt{1 + \frac{\Pi^2}{T_1^2 \lambda^2}} \left(2\pi a_0^2 r T_1 + 2 n_{\psi} \times \frac{1}{2} T_1 \frac{R^2}{r} n_\theta^2 \int_{\psi_i=m\pi}^{\psi_f=(m+1)\pi} \sin^2\psi d\psi \right) \nonumber \\
\sim  \sqrt{1 + \frac{\Pi^2}{T_1^2 \lambda^2}} \left(2\pi a_0^2 r T_1 +  2
n_\psi \times \frac{\pi}{4}T_1 \frac{R^2 n_\theta^2 }{r}\right) .
\end{eqnarray}
As the first term is simply the rest mass of the string in the Minkowski directions, the second term corresponds to the rest mass of $N=2n_\psi$ beads. 
Finally, setting $n_\psi \sim n_\theta \sim n$ explicitly, we have the following estimates for both the initial number of beads $N$ and the bead mass $M_b$,
%(3.12)
\begin{equation}
N \sim 2n \nonumber
\end{equation}
%(3.13)
\begin{equation}\label{eq:beadmassestimate}
M_b \sim \frac{\pi}{4} T_1 \sqrt{1 + \frac{\Pi^2}{T_1^2 \lambda^2}} \left(\frac{R^2 n^2}{ r} \right)
\end{equation}
Note that the bead mass is inversely proportional to $r$. This makes sense on the assumption that the total length of the string remains constant.
So if there is less string length involved in internal windings (resulting in less massive beads) more length is added to the ordinary four dimensional part of the loop. 
The converse is also true. Thus the first expression for the mass increases in magnitude as the second decreases and vice-versa. 
Note also that the $M_b$ is effectively quantised in terms of the number of windings present in the $\theta$-direction.

Let us now also consider the full expression for the bead mass. One can integrate the potential over the above range and the result is roughly the sum of the
bead mass and the rest mass associated with an unwound string (with $n=0$). We can then re-write the bead mass including all the higher order terms, with 
an appropriate normalisation to yield
\begin{equation}\label{eq:full_bead_mass}
M_b \sim 2 a_0^2 T_1 r \sqrt{1+\frac{\Pi^2}{T_1^2 \lambda^2}} \left({\rm EllipticE}\left(\frac{nR i}{a_0 r} \right)-\frac{\pi}{2} \right)
\end{equation}
which is valid for non-zero winding number $n$. One can easily check that in the limit where $nR<< a_0 r$ the mass reduces exactly to the one in (\ref{eq:beadmassestimate}).
%Section4%%%%%%%%%%%%%%%%%%%%%%%%%%%%%%%%%%%%%%%%%%%%%%%%%%%%%%%%%%%%%%%%
%%%%%%%%%%%%%%%%%%%%%%%%%%%%%%%%%%%%%%%%%%%%%%%
\section{Cosmological implications of necklace loops}\label{cosmology}
We now investigate the cosmological implications of necklace loops based on the assumption that they retain their necklace structure after formation. 
As we have already seen, the $\theta$-dependence of the lifting potential depends on the presence of windings in the $\phi$-direction ($n_\phi>0$) and its 
$\psi$-dependence requires the presence of windings in the $\theta$-direction ($n_\theta>0$). However since the $\phi$-direction is flat, windings 
along this direction are unstable. If these windings contract this effectively flattens the $\theta$-direction leaving the $\theta$-windings free to contract as well. 
This in turn flattens the $\psi$-direction and necklace structure disappears as the bead mass (which comes from extra-dimensional windings) 
is converted into the ordinary rest mass of a loop in Minkowski space. We
therefore see that, due to the presence of single flat direction in the $S^3$, 
the entire necklace structure of the loop is unstable and may unravel in time.

We will consider this possibility later in the present section where we will introduce a time-dependent model for the number of windings. For the 
moment we will we will consider $n \sim n_\psi \sim n_\theta \sim n_\phi$ to be roughly constant.
We can now use Matsuda's original assumption that the four-dimensional part of the loop loses mass-energy via the emission of gravitational radiation in Minkowski space 
(just like an ordinary cosmic string in four dimensions) but that the bead-mass which is formed from the winding of the string in internal space is unaffected by 
this process. 

To investigate the cosmological implications of necklace loops we must therefore modify equation (\ref{eq:totalpotential}) by inserting
a time-dependent radius $r(t,t_i)$ into the first term of the expansion (the loop mass in Minkowski space) and the initial 
radius $r(t_i)$ into the second term (the bead mass). The time dependent radius of a shrinking loop in warped Minkowski space is given by, 
%(4.1)
\begin{equation}
a_0 r(t,t_i) = a_0 (\alpha t_i - \Gamma \mathcal{G} T_1 (t - t_i)) 
\end{equation}
where $t_i$ is the time of loop formation. Note that $t \geq t_i$ is the cosmic time coordinate, $\Gamma$ is a measure of the rate of 
energy loss due to the emission of gravitational radiation, $\mathcal{G}$ is Newton's constant (which is determined by the volume of the $S^3$) 
and $0 < a_0 \alpha < a_0$ determines the characteristic initial loop radius as a fraction of the horizon, $d_H \sim a_0 ct$ The initial loop radius as a function of $t_i$ is then
%(4.2)
\begin{equation}\label{eq:initial_loop_radius}
a_0r(t_i) = a_0\alpha t_i. 
\end{equation}  
The bead mass now explicitly depends on the time of loop formation, $t_i$ via (\ref{eq:initial_loop_radius}). However we also expect the initial number of beads, $N$, 
will also depend in some way on $t_i$. We therefore need some way of estimating the initial number of windings in each direction $n = n(t_i)$.

Following Shellard and Avgoustidis \cite{Avgoustidis:2004zt}(and taking into account the warp factor $a_0$) we use the random walk regime to estimate the initial number of windings present in each angular 
direction, $n(t_i)$, so that\footnote{Although in general a random walk will not give rise microscopically to completely 
smooth windings, as described by the embedding ansatz, it should on average produce something similar from a macroscopic point of view. 
Any extra beads formed by microscopic structure would quickly annihilate one another if we assume that they are free to move around the 
\emph{shrinking} loop in a random walk. 
This random motion occurs when small sections of the string momentarily acquire enough energy to jump between adjacent minima in the effective potential, 
causing the bead to move from a four-dimensional perspective. In his original paper \cite{Matsuda:2006ju} Matsuda predicted that, based on the idea of a random walk, 
approximately $\sqrt{n(t_i)}$ beads/anti-beads created in pair formation would survive until late times. However we contest that this is valid 
only for a \emph{static} loop. For a shrinking loop it seems clear that all bead-anti-bead pairs will eventually collide and annihilate, 
at least if the minimum radius of the string ($\sim$ string thickness) is comparable to the initial spacing. 
This implies that only beads formed from \emph{net} windings will contribute to the mass of any PBH's/DM relics eventually created. 
Further discussion of this point is given in Section $6$.}
%(4.3)
\begin{equation}\label{eq:n}
n(t_i) = \frac{\sqrt{a_0\alpha \omega_l \epsilon_l t_i}}{R}
\end{equation}
where $0 < \omega_l < 1$ is the fraction of the total string length l (not to be confused with the angular momentum) 
contained in the windings and $\epsilon_l$ is the step length. This definition tells us that string networks will tend to form once the 
correlation length becomes larger than the scale of the internal dimensions. 

The parameter $\omega_l$ may be defined as the ratio of the string length in the internal dimensions to the
total length of the string in all dimensions. 
Following \cite{Avgoustidis:2004zt}, but using the notation defined earlier, the concise definition of $\omega_l$ is given by
\begin{equation} \label{eq:VOS}
\omega_l = \sqrt{\int \left(\frac{R^2 s'^2}{a_0^2 r^2 + R^2 s'^2}\right) \mathcal{V} d\sigma/ \int \mathcal{V} d\sigma}
\end{equation}
where $\mathcal{V}(\sigma)$ denotes the integrand in equation (\ref{eq:effectivepotential}). 
In order to simplify the above expression we then make the following approximations,
\begin{eqnarray}
\omega_l &\sim& \sqrt{\frac{\int R^2 s'^2 d\sigma }{\int (a_0^2 r^2 + R^2 s'^2)d\sigma} \int \mathcal{V} d\sigma/ \int \mathcal{V} d\sigma}
\nonumber\\
&\sim& \sqrt{\frac{n^2 R^2 \int (4 + \sin^2(n\sigma) + \sin^4(n\sigma)) d\sigma }{a_0^2 r^2 + n^2 R^2 \int (4 + \sin^2(n\sigma) + \sin^4(n\sigma)) d\sigma}}
\end{eqnarray}
where in the second step we have ignored the numerical coefficient in front of the $r^2$ term, 
and again used assumption that $n \sim n_{\psi} \sim n_{\theta} \sim n_{\phi}$. 
Finally we note that it is straightforward to show that
\begin{equation}
\int (4 + \sin^2(n\sigma) + \sin^4(n\sigma)) d\sigma \sim O(10)
\end{equation} 
for all possible values of $n(t_i)$. Therefore we see that $\omega_l$ can be well approximated by the following function
\begin{equation}\label{eq:omega}
\omega_l \sim \frac{nR}{\sqrt{a_0^2 r^2 + n^2 R^2}}
\end{equation}
up to various numerical factors which would have appeared in the above expression if the integral in (\ref{eq:VOS}) been performed in full. 
It is clear however, that $\omega_l$ has the correct functional dependence on both $n$ and $r$. 
This means that, whatever time-dependent models we use for $r(t_i)$, $n(t_i)$ and $\omega_l(t_i)$ - they must satisfy (\ref{eq:omega}) for consistency. 
Note that at this stage we are not imposing any additional conditions on $a_0 r$ and therefore $\omega_l$ is just a parameter of the theory.

If we then use (\ref{eq:initial_loop_radius}) as our model for $r(t_i)$ and substitute $\omega_l$ from (\ref{eq:omega}) into our 
previous expression for $n(t_i)$ (which also uses the approximation $r(t_i) \sim \alpha t_i$) we find a cubic equation in the variable $(nR)^2$,
%(4.5)
\begin{equation}
(nR)^6 + a_0^2(\alpha t_i)^2(nR)^4 - a_0^2 (\alpha t_i)^2 \epsilon_l^2 (nR)^2 =0
\end{equation} 
This has the trivial solution $(nR)^2=0$ (no windings) and the more physically relevant one,
%(4.6)
\begin{equation}\label{eq:nsolution}
(nR)^2 = \frac{a_0^2 (\alpha t_i)^2}{2}\left(-1 \pm \sqrt{1+\frac{4 \epsilon_l^2}{a_0^2 (\alpha t_i)^2}} \right)
\end{equation} 
although the reality of $n$ requires us to take the positive sign before the square root. 
The time dependence of $n(t_i)$ in the expression above is then consistent with the definition of $\omega_l$ in (\ref{eq:omega}).
  
Next we move on to consider to size of the step length $\epsilon_l$. 
The maximum velocity of the string in the compact dimensions is $c$ (where $c=1$ in natural units). 
which corresponds to a step length per unit time ($\Delta t = a_0^{-1}t_s = a_0^{-1} \sqrt{\alpha'}$) of $\epsilon_l = a_0^{-1}l_s 
= a_0^{-1} \sqrt{\alpha'}$.
\footnote{Here we have used the fact that the standard string tension $T_1 \sim \alpha'^{-1}$ is given by the fundamental string mass divided by the fundamental string length $T_1 = \frac{m_s}{l_s} \sim \frac{\sqrt{\alpha'}^{-1}}{\sqrt{\alpha'}} \sim \alpha'^{-1}$ together with the fact that the \emph{effective} tension of the string in the warped throat is $\tilde{T_1} \sim a_0^2 T_1$. We then treat this as the division of the 'warped' string mass $\tilde{m_s} \sim a_0 \sqrt{\alpha'}^{-1}$ by the 'warped' string scale $\tilde{l_s} \sim a_0^{-1} \sqrt{\alpha'}$ so that the velocity of light is given by $c = \frac{\tilde{l_s}}{\tilde{t_s}} = \frac{a_0^{-1} \sqrt{\alpha'}}{a_0^{-1} \sqrt{\alpha'}} \equiv \frac{l_s}{t_s} = \frac{\sqrt{\alpha'}^{-1}}{\sqrt{\alpha'}} = 1$}
However only the endpoints of the string at the horizon move at the speed of light. 
For two points on the string within the horizon, separated by a distance $d \sim \alpha c t$, the relative velocity between them is $v \sim \alpha c$
which corresponds to an effective step length of,
\begin{equation} \label{eq:epsilon}
\epsilon_l \sim \alpha a_0^{-1}\sqrt{\alpha'}  
\end{equation}
Note that this definition also implies that $n(t_i) \propto \alpha$ as we would expect. This ensures that the number of beads on a 
long string, which stretches across the entire horizon $n_{Hor}(t_i)$, is independent of $\alpha$. In fact 
the assumption of a constant step length $\epsilon_l$ at all points along the string is problematic, 
as this leads to a measure of $n_{Hor}(t_i)$ which is proportional to $1/\sqrt{\alpha}$.  

Using (\ref{eq:epsilon}) and (\ref{eq:nsolution}) the condition for bead formation, $n(t_i)^2 \geq 1$, is then equivalent to, 
\begin{equation}
t_i \geq \frac{R^2}{\alpha \sqrt{\alpha^2 \alpha' - a_0^2 R^2}}
\end{equation}
and reality of the above solution translates into the constraint
\begin{equation}\label{eq:primeconstraint}
a_0^2 \leq \frac{\alpha^2 \alpha'}{R^2}.
\end{equation}
Now strictly speaking, $a_0^2$ is fixed by the ratio of the fluxes arising 
in a full string compactification and is therefore highly sensitive to the magnitude of string coupling, the string scale and the flux parameters. 
In the non-compact case which we are considering we also require that $a_0^2 < 1$ in order to ensure that the solution is warped. 
Therefore the existence condition for beads appears to impose a strict upper bound on the value of $a_0^2$ which in turn imposes a strong
constraint on the deformation parameter of the conifold geometry.
We also note that if $a_0^2 = \frac{\alpha^2\alpha'}{R^2}$, this allows necklaces to form only at infinity, and so only the strict inequality 
in (\ref{eq:primeconstraint}) is physically meaningful.

One can now substitute the solution for $nR$ into the generalised mass function in (\ref{eq:full_bead_mass}). If we define the following function
\begin{equation}
G(t_i) = -1 + \sqrt{1+\frac{4 \alpha'}{a_0^4 t_i^2}}
\end{equation}
then we see that there is a remarkable cancellation of terms and we are left with
\begin{equation}
M_b \sim 2 a_0^2 T_1 \alpha t_i \sqrt{1+\frac{\Pi^2}{T_1^2 \lambda^2}} \left({\rm EllipticE}\left( i \sqrt{\frac{G(t_i)}{2}} \right)-\frac{\pi}{2} \right)
\end{equation}
which is not a monotonic function of time. 
For vanishingly small $t_i$ the mass is increasing until it reaches a maximum value, before decreasing monotonically.
This leads to the interesting possibility that one can have sustained PBH formation during a small window where the mass function is greater than the critical value required to produce a gravitational radius larger than the string width. This possibility is dealt with explicitly in section 5.4.

We also note that the explicit form of $\omega_l(t_i)$ is given by, 
\begin{equation}
\omega_l \sim  \frac{a_0^2 t_i}{2\sqrt{\alpha'}}\left(-1 + \sqrt{1+\frac{4 \alpha'}{a_0^4 t_i^2}}\right) = \frac{a_0^2 t_i}{2\sqrt{\alpha'}}G(t_i)
\end{equation}
which tends to unity as $t_i \rightarrow 0$ and zero as $t_i \rightarrow \infty$ as we would expect.

To aid visualisation in the discussions that follow we plot, in Figure 3, the 
functions $M_b(t_i)$, $n(t_i)$ and $\omega_l(t_i)$ and the inter-bead distance $d(t_i)$ for
 fixed values of our model parameters to illustrate the qualitative behaviour of each. 
\begin{figure}[htp]
 \begin{center}
 {\includegraphics[width=0.9\textwidth]{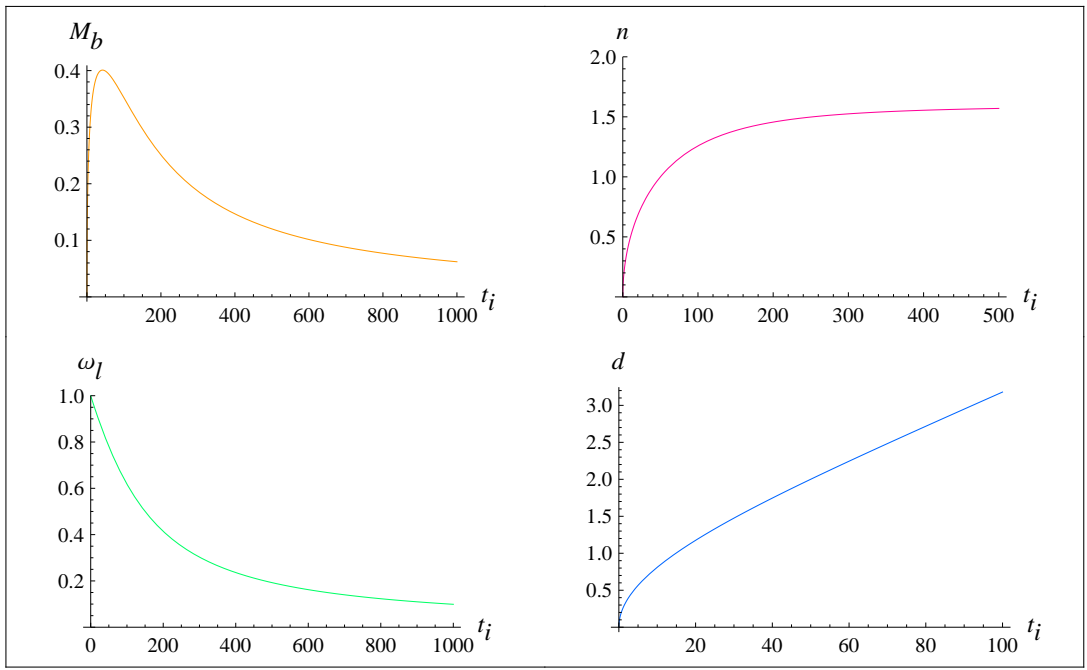}} 
  \caption{Plots of the mass of an individual bead $M_b$ as a function of time $ t_
i$, the number of beads $n(t_i)$, $\omega_l(t_i)$ and the average inter-bead distance $d(t_i)$.
In all plots model parameters have been set such that $a_0 = 0.1 $, $\alpha' =1$, $R=5$, 
$\alpha=0.8$ and $T_1 \sqrt{1 + \frac{ \Pi^2}{T_1^2 \lambda^2}} =
 1$.}
 \end{center}
\end{figure}
What is clear from the plots is that, assuming no gravitational emission, there are initially few beads with large mass. However as time increases, the number
density of beads gets larger and their corresponding mass decreases. 
The rate of decrease is in fact much larger than the increase in the number density. Therefore one expects that eventually the system will become 
dominated by a large number of (almost) massless beads. 
%%%%%%%%%%%%%%%%%%%%%
\subsection{Late time formation - the scaling regime}\label{section:late_time}
Using the fact that $nR$ is fixed by the parameters of the theory, let us consider the asymptotics to better understand the physics at early and late
epochs. Let us initially focus on the late time regime where
\begin{equation}
t_i >> \frac{2\sqrt{\alpha'}}{a_0^2 }
\end{equation}
which, using (\ref{eq:nsolution}), implies that the initial number of beads is constant
%(4.10)
\begin{equation} \label{eq:n_squared}
n^2 \sim \frac{\alpha^2 \alpha'}{a_0^2 R^2}.
\end{equation} 
Note that the condition for bead formation $n(t_i)^2 \geq 1$ simply reproduces (\ref{eq:primeconstraint}), ensuring the reality of $t_i^{(min)}$ -
the minimum time at which necklaces formation begins - and is consistent 
with our expectations.
We are then able to estimate the average inter-bead distance for necklaces formed at late times
%(4.11)
\begin{equation}
d(t_i) \sim \frac{a_0 r(t_i)}{N} \sim \frac{a_0^2 R  t_i}{2 \sqrt{\alpha'}} .
\end{equation} 
Identifying this with the correlation length along the string, $L(t_i)$, as mentioned in the previous section, we see that our solution 
corresponds to a scaling regime in which we may identify the scaling parameter $\gamma$ with the Klebanov-Strassler parameters, i.e.
%(4.12)
\begin{equation}\label{eq:bead_distance}
d(t_i) \sim L(t_i) \sim a_0 \gamma t_i    
\end{equation} 
implying that,
%(4.13)
\begin{equation}\label{eq:stringgamma}
\gamma \sim \frac{a_0 R }{2\sqrt{\alpha'}}   
\end{equation} 
Compare this with the result of field theory calculations, where the initial number of beads on a loop of radius $r(t_i)$ in the scaling regime is given by 
\cite{book},
%(4.14)
\begin{equation}
N(t_i) \sim \frac{r(t_i)}{L(t_i)} \sim \frac{\alpha}{\gamma}   
\end{equation} 
Hence we see that the condition for bead formation in the string picture (\ref{eq:primeconstraint}) 
is equivalent to the condition for bead formation in the scaling regime of the field theory, $\gamma \leq \frac{1}{2}\alpha$. 
Both conditions ensure that $N \geq 2$, implying that beads come in pairs. 
This is also consistent with the identification of the inter-bead distance with degenerate vacuum states, or correlation distance, 
along the string. Thus we see that a scaling solution arises naturally at late times in the string picture allowing us to 
identify various string parameters with their field theory counterparts. 

Using the condition on $\gamma$ and $\alpha$ which is consistent with bead formation 
i.e. $0 < \gamma \leq \frac{1}{2}\alpha$, this simply recovers our previous constraint on $a_0R$ (\ref{eq:primeconstraint}). 
However by setting $\alpha$ to the maximum value allowed by causality $\alpha = 1$ we may place a maximum upper bound on the radius of the $S^3$
%(1.8)
\begin{equation}
R \leq a_0^{-1} \sqrt{\alpha'}.
\end{equation}
This simply means that beads/windings will not form unless the radius of the $S^3$ is smaller than the warped string scale, 
irrespective of the value of $\alpha$. It also implies
\begin{equation}
a_0^2 \leq \frac{1}{bMg_s}
\end{equation}
in order for windings to form. Note also that setting $\epsilon_l \propto \alpha$ ensures that $n(t_i) \propto \alpha$ even if $\omega_l$ is  left  
as a free parameter as in \cite{Avgoustidis:2004zt}. 
Otherwise we have that $n(t_i) \propto \sqrt{\alpha}$ which implies that the number of windings per 
Hubble volume is proportional to $1/\sqrt{\alpha}$ which seems unphysical.

We now ask what happens to the time dependence of the mass function in the late time limit. The argument of the EllipticE term becomes small and 
therefore we may series expand the mass term to obtain
\begin{equation}
M_b \sim  T_1 \sqrt{1+\frac{\Pi^2}{T_1^2 \lambda^2}} \frac{\pi}{2}\frac{\alpha^2 \alpha'^{\frac{3}{2}}}{a_0^3 t_i} \left(1-\frac{3}{16} \frac{\alpha'}{a_0^4 t_i^2}+\ldots \right)
\end{equation}
and therefore we see that the mass evolves inversely proportional to $t_i$ and that $M_b \to 0$ as $t_i \to \infty$. This implies that any necklace structure
should disappear at sufficiently late times. In other words it is unlikely that any necklaces formed at late times would be distinguishable from
ordinary string loops. One should quantify this by noting that the mass is also inversely proportional to the scale of the warping, therefore for highly warped
throats the mass will be constant over a much larger time scale.
%%%%%%%%%%%%%%%%%%%%%%%%%%%%%%%%%%%%%%%%%%%%%%%%%%%%%%%%%%%%%%%%%%%%%%%%%%%%%%%%%%%%%%%%%%%%%%%%%%%%%%%%%%%%%%%%%%%%%%%%%%%%%%%%%%%%%%%%%%%%%%%%%%%%%%%%%%%%%%%%%%
\subsubsection{Early time formation}
Returning now to the cubic solution for $(nR)^2$, let us consider early time formation subject to the condition 
that $t_i \leq \frac{2\sqrt{\alpha'}}{a_0^2}$. This gives us
%(4.18)
\begin{equation}\label{eq:earlytimebound}
n^2(t_i) \sim \frac{\alpha^2 t_i \alpha'}{R^2} - \frac{1}{2} \left(\frac{a_0}{R}\right)^2 (\alpha t_i)^2 
\end{equation}
to second order, which also implies that,
%(4.19)
\begin{equation}
\omega_l(t_i) \sim 1 -\frac{1}{2\sqrt{\alpha'}} a_0^2 t_i
\end{equation}
from the definition of $\omega_l$. Clearly therefore $\omega_l \rightarrow 1$ as $t_i \rightarrow 0$, 
that is as $a_0 r(t_i) \rightarrow \mathcal{O}(nR)$. Utilising the bead-formation condition then allows us to place a bound on the bead formation time via
\begin{equation}
\frac{R^2}{\alpha \sqrt{\alpha^2 \alpha' - a_0^2 R^2}} \leq t_i \leq \frac{2\sqrt{\alpha'}}{a_0^2}.
\end{equation}
The average distance between beads, should necklaces begin to form\footnote{Note that we have assumed $t_i^{(min)} \sim \frac{R^2}{\alpha \sqrt{\alpha^2 \alpha' - a_0^2 R^2}} 
\leq \frac{2\sqrt{\alpha'}}{a_0^2}$ which is equivalent to assuming $a_0 R \leq 0.91 \alpha \sqrt{\alpha'}$}., may then be approximated by
%(4.20)
\begin{equation}\label{eq:early_bead_distance}
d(t_i) \sim \frac{a_0 R \sqrt{t_i}}{2 \alpha'^{1/4}} \left(1+\frac{a_0^2 t_i}{4 \sqrt{\alpha'}}+\ldots \right)
\end{equation}
which, unlike the the late time approximation, does not correspond to any known regime in the field theory if we continue to identify $d(t_i)$ 
with the correlation distance $L(t_i)$. However we note that the second term is sub-dominant at very early times when 
$t_i \leq \frac{4\sqrt{\alpha'}}{a_0^2}$ suggesting that it may be reasonable to keep only first order terms in the expansion
especially if $t_i << \frac{2\sqrt{\alpha'}}{a_0^2}$. 
Such a very early time approximation may correspond to a damping regime in the field theory picture. 
The above equation would then represent an intermediate regime where damped and scaling solutions can be joined together.
Assuming now that $t_i << \frac{2\sqrt{\alpha'}}{a_0^2 \alpha}$ and keeping only the first order term in (\ref{eq:earlytimebound}), 
we see that the condition for bead formation $n^2(t_i) \geq 1$ is then equivalent to, 
%(4.22)
\begin{equation}
t_i \geq \frac{R^2}{\alpha^2 \sqrt{\alpha'}}
\end{equation}
This is consistent with our previous estimates and shows that bead formation will occur in the very early time regime
\footnote{Note that the condition$t_i^{(min)} \sim \frac{R^2}{\alpha^2 \sqrt{\alpha'}} \leq \frac{2\sqrt{\alpha'}}{a_0^2}$ 
implies that $a_0 R \leq \sqrt{2} \alpha \sqrt{\alpha'}$ which is automatically satisfied by (\ref{eq:primeconstraint}).} only when 
$a_0^2 R^2 << \alpha^2 \alpha'$. With this in mind we see that the inter-bead distance 
is set by the leading factor in (\ref{eq:early_bead_distance}) which is consistent with the field theory picture 
at early times during the damping regime, where we expect the correlation distance $L(t_i)$ to be given by a power law solution of the form \cite{book},
%(4.24)
\begin{equation}\label{eq:correlation_distance_early}
L(t_i) \sim t_d^{\frac{1}{2}}t_i^{\frac{1}{2}}
\end{equation}
with $t_d$ corresponding to the characteristic damping time of small scale oscillations on the string. 

However a damping regime is usually obtained by considering collisions of the string with an external plasma. 
In this case the damping term comes from the internal dynamics of the model, suggesting that the inertia of the beads (when $M_b$ is large) 
is sufficient to cause the correlation length to scale as $L(t_i) \sim \sqrt{t_i}$ at very early times.

Furthermore unless the warping is extraordinarily large, the time-scales over which this effect takes place are likely to be insignificant in cosmological terms. 
As a future amendment to the current work therefore it will be useful to impose an external damping regime and to study the effect of the interaction of the 
windings with the external plasma. Naively we may expect collisions of the string with particles in the compact space to inhibit the formation of windings 
resulting in the delayed on-set of a scaling regime. This to is likely to mirror the field theory case, though further investigation is needed to 
establish whether the correlation distance scales according to (\ref{eq:correlation_distance_early}).  

Once again, for the sake of completeness, we consider how the mass function changes as a function of time in this epoch. The elliptic integral 
is actually divergent in this limit, however we 
can consider the leading order divergence which will dominate the spectrum. The resulting expression for the mass function will then behave like
\begin{equation}
M_b \sim 2 T_1\sqrt{a_0 \alpha'^{1/4}}\sqrt{1+\frac{\Pi^2}{T_1^2 \lambda^2}}(\alpha t_i)^{3/4} 
\end{equation}
which one can see will tend to zero as $t_i$ goes to zero. This is within the regime where we expect PBH formation to occur, since the mass of the 
necklace plus beads is steadily increasing (as a function of time) during this regime. 
%Subsection4.3%%%%%%%%%%%%%%%%%%%%%%%%%%%%%%%%%%%%%%%%%%%%%%%%%%%%%%%%%%%
%%%%%%%%%%%%%%%%%%%%%%%%%%%%%%%%%%%%%%%%%%%%%%%
\subsection{PBH formation}
We now calculate the contribution to the PBH mass spectrum from collapsing necklace loops, based on the assumption that loops which 
chop off from the string network retain their necklace structure indefinitely. 
We will initially take the number of beads per loop to be constant from the time of formation.

The minimum radius to which a contracting loop may shrink, $\delta_{l}$, is limited by the string width - which we assume to be comparable to the 
inverse of the symmetry breaking scale $\eta_s$ \cite{Matsuda:2006ju, Matsuda:2005ez}, i.e. $\delta_{l} \sim \eta_s^{-1}$ 
\footnote{It can be shown that, for Abelian field theories, the tension of a string is proportional to the 
square of the symmetry breaking scale, $\mu \sim \eta^2$. Our Abelian string with the same assumption gives $\eta_s \sim \sqrt{\alpha'}^{-1}$  
However we will leave $\eta_s$ as a free parameter in the discussion that follows.}.
The condition for gravitational collapse then becomes
%(4.37)
\begin{equation}\label{eq:gravitational_collapse}
R_S > \eta_s^{-1}
\end{equation}
where $R_S$ is the Schwarzschild radius of the loop. We may estimate the 
Schwarzschild radius of a necklace loop using the spherically symmetric approximation
%(4.38)
\begin{equation}
R_S(t_i) \sim 2 \mathcal{G}M_T(t_i)
\end{equation} 
where $\mathcal{G}$ is the modified Newton's constant and $M_T(t_i)$ is the total mass of the necklace. A necklace formed at time $t_i$ will therefore collapse to form a PBH if, 
\begin{equation} \label{eq:pbhcondn}
M_T(t_i) \geq \frac{\eta_s^{-1}}{2\mathcal{G}} 
\end{equation} 
Assuming that $\delta_l$ is small, the bead mass will provide the dominant contribution to $M_T(t_i)$ so that
\begin{eqnarray}\label{eq:pbhexact}
M_T(t_i) &\sim&  N(t_i)M_b(t_i)  \nonumber \\
&\sim&  \frac{8 T_1 a_0^3}{\sqrt{2}R}\sqrt{1+\frac{\Pi^2}{T_1^2 \lambda^2}} (\alpha t_i)^2 \sqrt{G(t_i)} \left({\rm EllipticE}\left(i\sqrt{\frac{G(t_i)}{2}} \right)-\frac{\pi}{2} \right)
\end{eqnarray}
which is not a monotonic function.
We can approximate the solution at early and late time regimes respectively using the following
\begin{eqnarray} \label{eq:early_late_approx}
M_T(t_i) &\sim& T_1 \sqrt{1+\frac{\Pi^2}{T_1^2 \lambda^2}}\frac{\pi}{4}\frac{\alpha^2 \sqrt{\alpha'}}{R} a_0 t_i + \ldots \nonumber \\
& \sim& T_1 \sqrt{1+\frac{\Pi^2}{T_1^2 \lambda^2}}\frac{\pi}{2}\frac{\alpha^2 \alpha'^{\frac{3}{2}}}{a_0^3 R} \frac{1}{t_i} \left(1-\frac{3\alpha'}{16 a_0^4 t_i^2} \right)
\end{eqnarray}
Here we note that $M_T$ is linearly increasing with $a_0 t_i$ at very early times, whilst in the late time epoch we find that the scaling is with $1/(a_0^3 t_i)$. Thus
the sensitivity to the warp factor is most pronounced at late times, since a vanishingly small value of $a_0$ means that the total mass remains larger over
a wider time-scale. Ultimately however, the total mass will tend to zero asymptotically.
The complete mass function $M_T(t_i)$, using the definition of the elliptic integral, is sketched in Figure 4 to illustrate the time dependence.
In Figure 5 this is shown together with the early and late time time approximations. 
\begin{figure}[htp]
 \begin{center}
  {\includegraphics[width=0.6\textwidth]{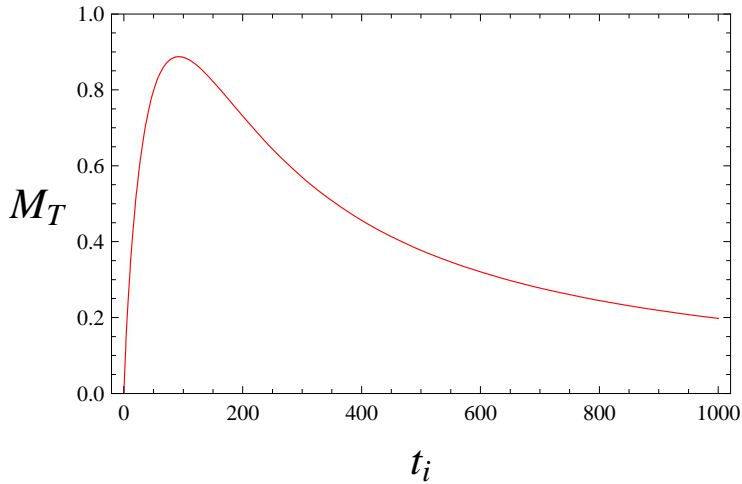}} 
  \caption{The total mass $M_T$ contained within the beads of a necklace 
as a function of $t_i$. 
The model parameters have been fixed so that $a_0 =0.1$, $\alpha=0.8$, $  \alpha' = 1$, 
 $T_1 \sqrt{1 + \frac{\Pi^2}{T_1^2 \lambda^2}} = 1$ and $R=5$.}
 \end{center}
\end{figure}
%Figures 5
\begin{figure}[htp]
 \begin{center}
  {\includegraphics[width=0.9\textwidth]{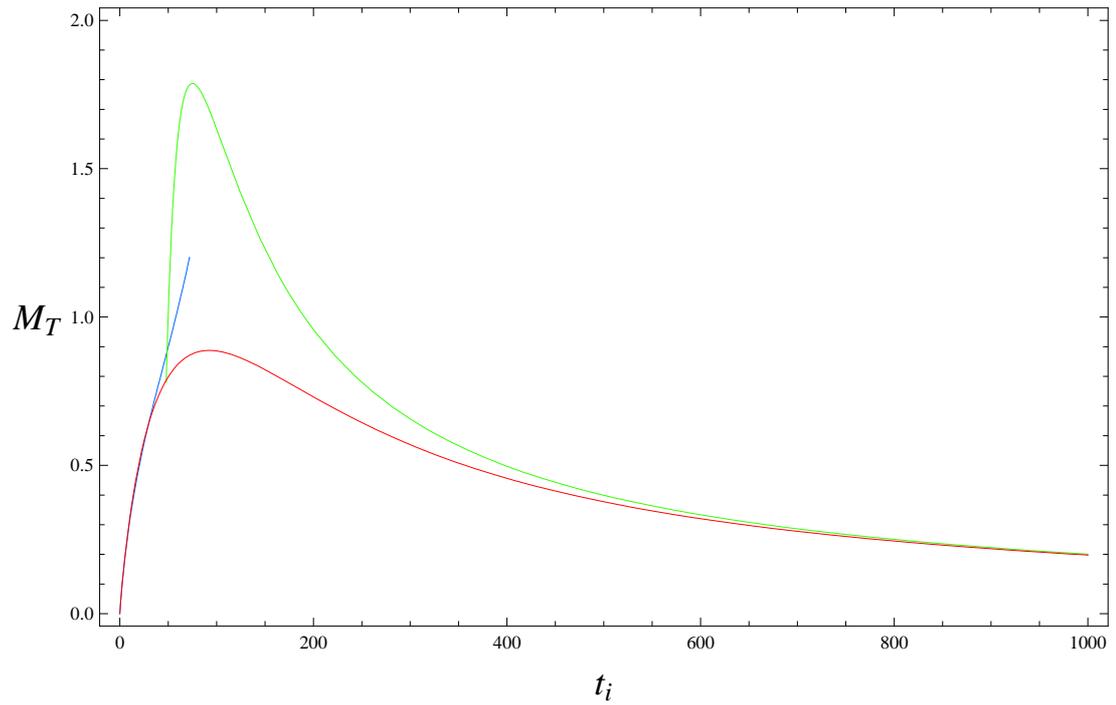}} 
  \caption{The total mass $M_T$ contained within the beads of a necklace 
as a function of $t_i$, together with the early and late time approximations given in equation
 $(\ref{eq:early_late_approx})$. 
The same parameter choices were made as in Fig. 4. 
The blue (green) curves are the early (late) time approximations respectively. Note that the late time
approximation seems to over-estimate the maximum mass.}
 \end{center}
\end{figure}
Without precise values for the model parameters it is difficult to determine the accuracy (as a percentage estimate) of either the early or late time 
approximations. However it is interesting to note that the late time approximation also qualitatively 
captures the behaviour of $M_T(t_i)$ in the early time regime. In fact the peaks of the two functions are in approximately the same position, 
although the peak of the approximation is roughly twice as high as the peak of the true curve. 

However it is also clear that the peak of the full mass function, which is potentially the region of greatest interest with respect to the formation of 
PBH's, lies between the regions in which either the early or late time approximations are valid. As we shall see, in order to calculate the PBH spectrum 
analytically in terms of our model parameters, we must integrate over the region of the curve (\ref{eq:pbhexact}) 
which satisfies (\ref{eq:pbhcondn}). This is not possible in general, as any such region (if it exists) will certainly include the peak itself. 
We must therefore find some way of approximating the elliptic integral in this crucial region. 

Whilst this can be done in many ways, we will use the following method. To begin with, we estimate the maximum height of the full mass function
by utilising the fact that the peak of the late time approximation lies at roughly the same value of $t_i$. 
Differentiating the second equation in (\ref{eq:early_late_approx}) and solving the resulting expression $dM_T/dt_i = 0$ gives,
\begin{equation}
t_i^{(peak)} \sim \frac{3}{4a_0^2}
\end{equation}  
Substituting this back into (\ref{eq:pbhexact}) then gives,
\begin{equation}
M_T^{(peak)} = M_T(t_i^{(peak)}) \sim 5.47 \ T_1   \sqrt{1+\frac{\Pi^2}{T_1^2 \lambda^2}} \frac{\alpha^2 \alpha'^{\frac{3}{2}}}{8a_0 R}
\end{equation}  
We then note that substituting the full expression for $n(t_i)$ (\ref{eq:nsolution}) into (\ref{eq:beadmassestimate}) and using 
$M_T(t_i) \sim N(t_i)M_b(t_i)$ from (\ref{eq:pbhexact}) also gives a function which shows the same qualitative behaviour as the true $M_T(t_i)$
curve (as we would expect). Taking this approach corresponds to expanding the elliptic integral only to first order in $n(t_i)$, but keeping the 
full time dependence of this function (which is valid even at early times) as opposed to keeping higher order terms in the expansion of 
the elliptic integral and using the late time approximation $n(t_i) \sim \alpha \sqrt{\alpha'}/a_0 R$ as in (\ref{eq:early_late_approx}). 
The resulting expression for $M_T(t_i)$ is,
\begin{equation} \label{eq:first_order_in_n}
M_T(t_i) \sim \frac{\pi}{\sqrt{32}} T_1 \sqrt{1+\frac{\Pi^2}{T_1^2 \lambda^2}} \frac{a_0^3 (\alpha t_i)^2}{R} \left(-1 + \sqrt{1 + \frac{4\alpha'}{a_0^4 t_i^2}}\right)^{\frac{3}{2}} 
\end{equation}
Although this is still highly inaccurate within the region of the peak, a function \emph{of this form} may be used to capture the behaviour of the 
full expression right down to all but the earliest times (where the early time expansion above must again be used).\footnote{It is more useful in this 
respect to keep the full time dependence in $n(t_i)$ while expanding to only first order in $n$ than it is to expand to two or more orders in the
 argument of the EllipticE for large $t_i$.}. We may then fit a curve of the form, 
\begin{equation} \label{eq:MT_A_C}
M_{Tapprox}(t_i) \sim A \ T_1 \sqrt{1+\frac{\Pi^2}{T_1^2 \lambda^2}} \frac{a_0^3 (\alpha t_i)^2}{R} \left(-1 + \sqrt{1 + \frac{C\alpha'}{a_0^4 t_i^2}}\right)^{\frac{3}{2}} 
\end{equation}
where $A$ and $C$ are free parameters, to the true $M_T(t_i)$ curve by demanding the approximate function satisfies;
\begin{itemize}
\item to pass through the point $(t_i^{(peak)},M_T^{(peak)}) = \left(\frac{3}{4a_0^2},5.47 \ T_1 \sqrt{1+\frac{\Pi^2}{T_1^2 \lambda^2}} \frac{\alpha^2 \alpha'^{\frac{3}{2}}}{8 a_0 R} \right)$ and, 
\item to have the same asymptotic behaviour as as the full elliptic integral. This is equivalent to the requirement that $A = \frac{\pi}{\sqrt{32}}C^{-\frac{3}{2}}$.
\end{itemize} 
These two condition are sufficient to fix $A$ and $C$ uniquely giving,
\begin{eqnarray} \label{eq:AandC_values}
A &\rightarrow& 0.368657 \nonumber\\
C &\rightarrow& 5.25648
\end{eqnarray}
The resulting fit is remarkably good both at late times and in the vicinity of the peak and is shown for a selection of parameter values in 
Figure 6 below. Only at very early times does the approximation appear to break down, which becomes clear if we 'zoom in' 
near $t_i=0$ as shown in Figure 7. 
However, if we \emph{are} required to integrate over a time range that includes a region in 
which $M_{Tapprox}(t_i)$ given in (\ref{eq:MT_A_C}) becomes invalid, we may always split the resulting integral into two parts. 
It would then be necessary to integrate, using the early time approximation in (\ref{eq:early_late_approx}) and expanded to arbitrary order, 
over some range $0 \leq t_i \leq t_{E}$ (where $t_{E}$ marks the time at which 
the approximation starts to break down) -  whilst using the expression (\ref{eq:MT_A_C})/(\ref{eq:AandC_values}) 
over the remainder of the range $t_{E} \leq t_i \leq t_{F}$.
    
%Figure 6
\begin{figure}[htp]
 \begin{center}
  {\includegraphics[width=0.9\textwidth]{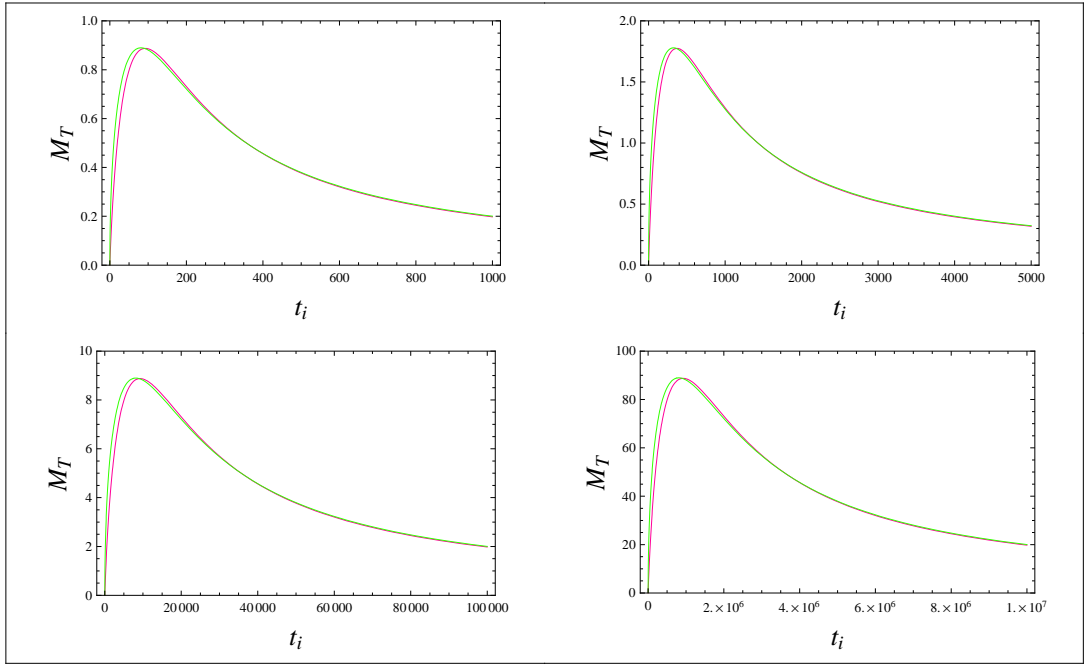}} 
  \caption{ Plot of the exact Necklace mass $M_T$ (red curve) and the approximate fit function $M_{Tapprox}$ (green curve)as a function of $t_i$ for 
  $a_0 = 0.1, 0.05,0.01,0.001$ (top left to bottom right). In all the plots $\alpha' =1$, $R=5$, $\alpha=0.8$ and $T_1 \sqrt{1 + \frac{ \Pi^2}{T_1^2 \lambda^2}} =
 1$.}
   \end{center}
\end{figure}
%Figures 7
\begin{figure}[htp]
 \begin{center}
  {\includegraphics[width=0.9\textwidth]{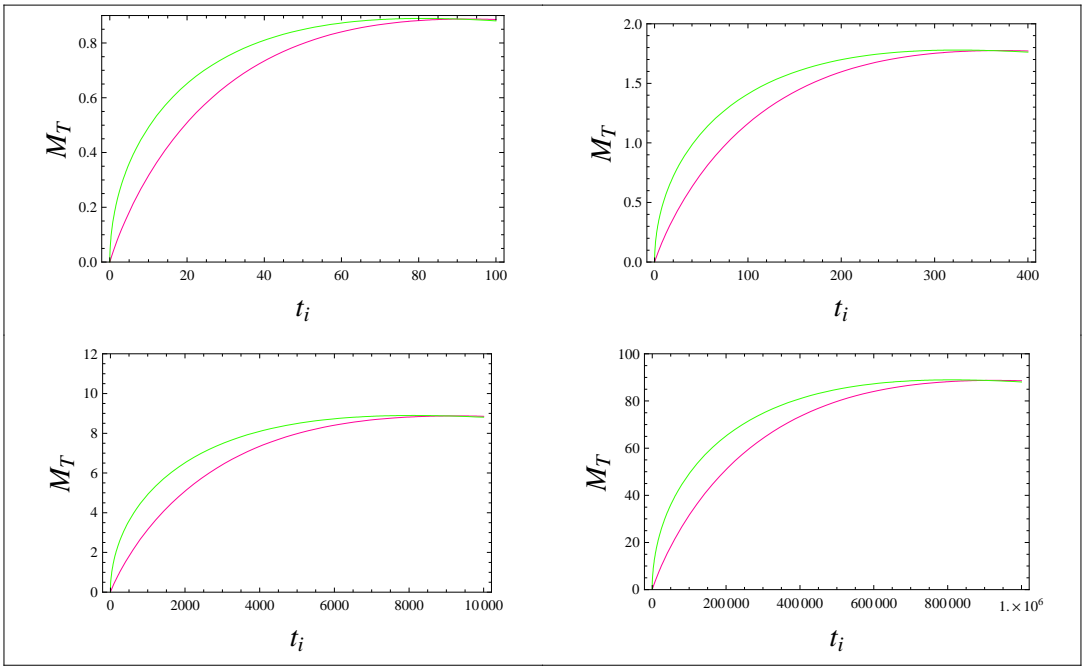}} 
  \caption{Zoom in of the four plots shown in Figure 6 clearly exhibiting the early time breakdown of the fit function $M_{Tapprox}(t_i)$  }
   \end{center}
\end{figure}
Starting with (\ref{eq:gravitational_collapse}) we may now estimate the range of time over which necklaces will form, 
that will eventually collapse to produce black holes i.e necklaces with sufficiently high mass contained in their beads to produce a Schwarzschild
radius greater than the string width. From our previous discussion we see that the mass function peaks at earlier times, therefore 
favouring PBH formation in this regime - that is, from the collapse of small necklace loops with a relatively low number of high mass beads.

In fact it is clear from $N(t_i)$, $M_b(t_i)$ in Figure 3, and especially from the plot $M_T(t_i) \sim N(t_i)M_b(t_i)$ 
in Figure 4, that the increase in bead number at late times is \emph{insufficient} to compensate for the reduction in the mass of individual beads. 
Similarly at very early times, there are an insufficient number of extra-dimensional windings to produce enough beads to create a large Schwarzschild radius, 
even though the individual bead mass is high. This results in the formation of a window in the early universe, during which 
necklaces suitable for the formation of PBH's are produced.

The limits of this window, let us call them $t_{PBH}^{\pm}$, may be estimated by solving equation (\ref{eq:pbhcondn})
 with the appropriate expression for $M_T(t_i)$. Let us now assume that the energy scale of the symmetry breaking process which gives rise to the $(p,q)$ string network is equal to the \emph{un-warped} string energy scale,
\begin{equation}
\eta_s \sim \sqrt{\alpha'}^{-1}
\end{equation}
and consider the limits which arise from inserting (\ref{eq:MT_A_C}) with the values of A and C given above. 
The resulting time bounds can be solved for analytically but yield rather complicated expressions which we do not repeat here.

These analytic expressions  are valid so long as the equation(\ref{eq:MT_A_C}) provides a relatively good approximation to the true $M_T(t_i)$ curve. 
However the validity of the lower bound, $t_{PBH}^{-}$, should always be checked (for a given set of parameter values) by plotting the full 
mass function together with the PBH formation bound,
\begin{eqnarray}
\frac{\eta_s^{-1}}{2 \mathcal{G}} \sim \frac{R^3 (\alpha')^2}{2} \frac{1}{G_{(10)}}
\end{eqnarray}
where we have used the fact that the effective four-dimensional gravitational coupling is 
related to the ten dimensional one through the relation
\begin{eqnarray}\label{eq:G_definition}
\mathcal{G} = \frac{G_{(10)}}{\rm{Vol}(X_6)},
\end{eqnarray}
where $X_6$ is the internal manifold. We can approximate the coupling via
\begin{equation}
\mathcal{G} \sim \frac{G_{(10)}}{R^3 (\alpha')^{3/2}}
\end{equation}
although we will typically assume that $G_{(10)} \sim \mathcal{O}(1)$ (in units where $\alpha' =1$) for simplicity.
In the event that this forms a poor estimate of the true $t_{PBH}^{-}$ it is necessary to expand the early time 
approximation given in (\ref{eq:early_late_approx}) to as many orders as required to reach an accurate value and 
to re-solve (\ref{eq:pbhcondn}). 

For the sake of completeness we note that the early time expansion given to first order, as 
in (\ref{eq:early_late_approx}), remains reasonably accurate up to,
\begin{equation} \label{eq:t_E}
t_{E} \sim \frac{\sqrt{\alpha'}}{2a_0^2}
\end{equation} 
beyond which the function (\ref{eq:MT_A_C}) is undoubtedly valid. The resulting estimate 
of $t_{PBH}^{-}$ (where $t_{PBH}^{-} < t_{E} \sim \frac{\sqrt{\alpha'}}{2a_o^2}$) is,
\begin{equation}
t_{PBH}^{-} \sim  \frac{R^4}{8 a_0 {\alpha'}^{5/2} \alpha^2 T_1\sqrt{1+\frac{\Pi^2}{T_1^2 \lambda^2}}}
\end{equation} 
In order to be safe therefore we may choose to split up any integral involving the the necklace mass 
$M_T(t_i)$ into two pieces: 
the first using the early time approximation (\ref{eq:early_late_approx}) between the limits $t_{PBH}^{-} \leq t_i \leq t_{E}$
and the 
second part using (\ref{eq:MT_A_C}) between $t_{E} \leq t_i \leq t_{PBH}^{+}$ where $t_{E}$ is given by (\ref{eq:t_E}).

An estimate for $t_{PBH}^{+}$ can also be obtained using the late time expansion for $M_T(t_i)$ and this yields
 the simple expression:-
\begin{equation}
t_{PBH}^{+} \sim \frac{\pi {\alpha'}^{7/2}\alpha^2 T_1 \sqrt{1+\frac{\Pi^2}{T_1^2 \lambda^2}}}{a_0^3 R^4}
\end{equation}
The next step towards calculating the necklace contribution to the present day PBH spectrum is to estimate 
the time at which a collapsing loop reaches its 
minimum radius $\delta_l \sim \eta_s^{-1} \sim \sqrt{\alpha'}$. This depends on the time the loop formed $t_i$ via 
\begin{equation}\label{eq:loop_time}
t'(t_i) \sim \left(1 + \frac{\alpha}{\Gamma \mathcal{G} T_1} \right) t_i - \frac{\eta_s^{-1}}{\Gamma \mathcal{G} T_1}. 
\end{equation}
We expect, from our approximations, that $\delta_l$ will be small and therefore the second term will be sub-dominant. This 
implies that the corresponding time range over which PBH's actually form from the collapse of these loops is
given by ${t'}_{PBH}^{-} \leq t'(t_i) \leq  {t'}_{PBH}^{+}$ where
\begin{equation}
{t'}_{PBH}^{\pm} \sim \left(1 + \frac{\alpha}{\Gamma \mathcal{G} T_1} \right){t}_{PBH}^{\pm}  .
\end{equation}
The initial mass of a black hole forming at $t'(t_i)$ is therefore equal to $M_T(t_i)$,
 which can be compared with the mass of black holed
formed from ordinary string loops - $M_{PBH}(t_i) \sim \mu \alpha t_i$, 
which yields the present day mass spectrum \cite{Wichoski:1998ev}
\begin{equation} \label{eq:n_PBH}
\frac{dn_{PBH}(M)}{dM} \propto M^{-2.5}.
\end{equation}
Clearly the mass spectrum for black holes formed via the collapse of necklaces is far more 
complicated, although the relevant calculation of the present
day spectrum proceeds in a similar fashion.
Firstly we identify
\begin{equation}
M_{PBH}(t'(t_i)) = M_{T}(t_i)
\end{equation}
Assuming also that the PBH formation window lies in the radiation dominated epoch, the number density of string loops with initial length 
$r(t_i) = \alpha t_i$ which chop off from the network at time $t_i$ is 
given by, 
%(4.52)
\begin{equation} \label{eq:loopdensity}
n(r(t_i),t_i) \sim \frac{\nu_r}{t_i^{\frac{3}{2}}(r(t_i))^{\frac{3}{2}}} = \frac{\nu_r}{\alpha^{\frac{3}{2}}t_i^3} 
\end{equation}
where $\nu_r$ is the number of long strings per Hubble volume. 
%We note that it appears consistent in the warped geometry
 %to account for the warping of both space \emph{and} time by introducing the transformations 
 %$t_i \rightarrow a_0 t_i$, $r(t_i) \rightarrow a_0 r(t_i)$ into the expression above, yielding an additional 
 %factor of $a_0^{-3}$. However, it is our belief that this effect is already accounted for via the
  %derivation of $\nu_r$ in terms of the warped throat model parameters. In fact such an explanation 
  %gives a nice interpretation to the formula (\ref{eq:nur1}) 
  %which is then seen as the product of three terms: a "warping term" which accounts 
  %for the back-reaction on the large dimensions $a_0^{-3}$, a term accounting for 
  %extra-dimensional effects $(\sqrt{\alpha'}/R)^{3}$ and a standard Lorentz factor $\sim g$. 
Now in field theoretic models,
\begin{equation}\label{eq:nur}
\nu_r = g \gamma^{-3} \tilde{c}
\end{equation}
where $g$ is a Lorentz factor, $\gamma$ gives the correlation length as a fraction of the horizon 
(which we previously identified with the string theory parameters $\gamma \sim \frac{a_0 R}{2 \sqrt{\alpha'}}$ (\ref{eq:stringgamma})) and $\tilde{c}$ is the 
loop production parameter. For ordinary four-dimensional strings $\tilde{c}$ is extracted from 
simulations and is of order unity. For higher dimensional strings $\tilde{c}$ is suppressed by a 
factor $P$ given by,
%\label{eq:P1}
\label{eq:P1}
\begin{equation}
P \sim \left(\frac{\delta}{R}\right)^{d-3}
\end{equation}
where $\delta$ is the effective string thickness and $d$ is the number of
 \emph{spatial} dimensions. In our model $\delta \sim \eta_s^{-1} \sim \sqrt{\alpha'}$ and $d=6$ giving,
%\label{eq:P2}
\label{eq:P2}
\begin{equation}
P \sim \left(\frac{\sqrt{\alpha'}}{R}\right)^{3}
\end{equation}
and hence,
\begin{equation}\label{eq:nur1}
\nu_r \sim \frac{8g}{a_0^3}\left(\frac{\sqrt{\alpha'}}{R}\right)^6
\end{equation}
In higher dimensional theories the scaling parameter $\gamma$ (which determines the correlation length) is
also expected to be suppressed by a factor of $P$ during the scaling regime, but small scale structure on the
strings is likely to lead to weaker $P$-dependence, $P_{\rm{eff}} = f(P)$ \cite{Avgoustidis:2005vm, Avgoustidis:2004zt}.
However as the correlation length has already been determined directly 
in terms of the model parameters it is likely that this effect has already been accounted for. 
We also note that it appears consistent in the warped geometry to account for the warping of both
space \emph{and} time by introducing the transformations $t_i \rightarrow a_0 t_i$, $r(t_i) \rightarrow a_0 r(t_i)$
into the expression (\ref{eq:loopdensity}) above, yielding an additional factor of $a_0^{-3}$. However it is our belief that
this effect is already accounted for via the derivation of $\nu_r$ in 
terms of the warped throat model parameters. In fact such an explanation gives a nice interpretation to the formula (\ref{eq:nur1}) which is then 
seen as the product of three terms: a "warping term" which accounts for the back-reaction on the large dimensions $a_0^{-3}$, a term accounting for 
extra-dimensional effects $(\sqrt{\alpha'}/R)^{3}$ and a standard Lorentz factor $\sim g$.

The formation rate of black holes at $t'(t_i)$ 
is then equivalent to (minus) the rate of necklace formation at $t_i$,
%(4.53)
\begin{equation}
\frac{dn_{PBH}(t'(t_i))}{dt'(t_i)} = -\frac{dn(r(t_i),t_i)}{dt_i} \sim \frac{3\nu_r}{\alpha^{\frac{3}{2}} t_i^4} 
\end{equation}
and finally we may use the the fact that
%(4.54)
\begin{eqnarray}
\frac{dn_{PBH}(t'(t_i))}{dM_{PBH}(t'(t_i))} &=& \frac{dn_{PBH}(t'(t_i))}{dt'(t_i)} \times \frac{dt'(t_i)}{dM_{PBH}(t'(t_i))} \nonumber\\
&\sim& \frac{3 \nu_r}{ \alpha^{\frac{3}{2}} t_i^4} \times \left( \frac{dM_T(t_i)}{dt_i} \right)^{-1}
\end{eqnarray}
to find the contribution to the PBH mass-spectrum from collapsing necklaces by substituting for $t'(t_i)$  and red-shifting to the current epoch. 

Of more interest for us is the calculation of the total contribution of the spectrum to the fraction of the critical density of the universe in the current 
epoch, $\Omega_{PBH}(t_0)$. The standard formula for $\Omega_{PBH}(t_0)$ from collapsing cosmic string loops is \cite{Matsuda:2005ez}
%(4.55)
\begin{equation}
\Omega_{PBH}(t_0) = \frac{1}{\rho_{crit}(t_0)} \int_{max(t_c,t_*)}^{t_0} dt \frac{dn_{PBH}}{dt} M(t,t_0)
\end{equation}
where $t_{*}$ is the formation time of a black hole with mass $M_{*} \approx 4.4 \times 10^{14} h^{-0.3}$ gm ($\equiv 10^{20}$ in Planck units), 
whose lifetime would be the present age of the universe\footnote{ Note that this value was calculated in a four dimensional FRW model 
using the standard value of $G = 6.67 \times 10^{-11} N m^2 kg^{-2}$. But in an a higher dimensional model, gravity is expected to become 
much stronger on very small scales resulting in a significantly higher rate of Hawking evaporation for the smallest PBH's. 
However we will neglect such small corrections.}. $M(t,t_0)$ is the current mass of a black hole that formed at a time $t$, 
and $t_c$ is the time at which loops first begin to form. 

Again assuming that most of the loop production occurs in the radiation dominated era, the rate of black hole formation is then given by
%(4.56)
\begin{equation}
\frac{dn_{PBH}}{dt} = \frac{3\nu_r f}{\alpha^{\frac{3}{2}} t^4} \frac{a(t)^3}{a(t_0)^3}
\end{equation} 
where $f$ is the fraction of loops which collapse on the first oscillation, which is expected to be small. 
It is typical to neglect the effect of Hawking radiation by making the approximation $M(t,t_0) \sim M(t)$, which in the standard case ($M(t) \sim \mu G t$) 
has been shown to make a difference of less than $6$ per cent to the final value of $\Omega_{PBH}(t_0)$ \footnote{Though in our model the difference may be substantially 
greater and is therefore something which should be checked for completeness.} \cite{Wichoski:1998ev}.

Adjusting the standard calculations to account for the effect of necklace formation, 
we expect there to be contributions to $\Omega_{PBH}(t_0)$ from two qualitatively different sources:
Firstly we expect to find a spike in the formation of PBH's in the very early universe due to the gravitational collapse of necklace loops 
which shrink to their minimum radius $\delta_l \sim \eta_s^{-1}$ within a time $t' \sim \left(1 + \frac{a_0 \alpha}{\Gamma \mathcal{G} T_1} \right) t_i$, 
although on cosmological time scales we may simply assume $t' \sim t_i$. 
The total mass of all the beads contained in these loops is large, and is therefore the dominant contribution to the initial mass of the black hole. 
This corresponds to the first term in (\ref{eq:omega_eqn}) below.

The second contribution to $\Omega_{PBH}(t_0)$ comes from loops which collapse \emph{before} shrinking to their minimum size,
by adopting a sufficiently compact and spherically symmetric configuration on their first oscillation \cite{book}. 
In principle this process is continuous through the history of the universe, although at late times we may neglect the contribution 
of the beads to the masses of black holes formed in this way - leaving just the first term in the second integral. At early times 
the beads must be included and therefore both terms become important. 

The expression for the (approximate) contribution of PBH's formed from collapsing loops to 
the current mass-density of the universe is therefore
%(4.57)
\begin{eqnarray}\label{eq:omega_eqn}
\Omega_{PBH}(t_0) \approx \frac{1}{\rho_{crit}(t_0)} \int_{t_*^-}^{t_*^+} dt_i \frac{3\nu_r}{ \alpha^{\frac{3}{2}}t_i^4}\frac{a(t_i)^3}{a(t_0)^3} M_T(t_i) \times \left(\frac{t_{*}^{+}}{t_r}\right)^{\frac{1}{2}} \left(\frac{t_r}{t_0}\right)^{\frac{3}{2}} \nonumber\\
+ \frac{1}{\rho_{crit}(t_0)} \int_{t_*^+}^{t_0} dt_i 
\left(\frac{6\pi \nu_r f T_1 \sqrt{1 + \frac{\Pi^2}{T_1^2 \lambda^2}}}{ \alpha^{\frac{1}{2}} t_i^3} + 
\frac{3 \nu_r f M_T(t_i)}{ \alpha^{\frac{3}{2}} t_i^4} \right)  \frac{a(t_i)^3}{a(t_0)^3} 
\end{eqnarray}
where $t_*^{\pm}$ are the times between which the necklace mass exceeds $M_*$ and $t_r$ marks the end of the radiation dominated era. 
In practice however the factor $f \sim 10^{-20}$ multiplying the integral on the second line indicates that (without fine tuning of the parameters) 
by far the largest contribution will come from the first integral between $t_*^-$ and $t_*^+$. 
In other words we expect the necklace-specific channel to dominate the production of PBH's and therefore choose to neglect the latter two terms. 

The two factors in large brackets outside the first integral account for the red-shifting from the \emph{end} of PBH production (with $M_{PBH} \geq M_*$) 
from necklace collapse to the present day. For our purposes it is convenient to use the parameterisation 
\begin{equation}
\rho_{crit}(t_0) \sim 3 H_0^2 M_p^2, \hspace{0.5cm} t_0 \sim \frac{2}{3} H_0^{-1}.
\end{equation}
%In Planck units $t_0 \sim 10^{60}$, $t_r \sim 10^{55}$ and $t_{*}^{+} \sim 10^{60}$ so that the total factor multiplying the integral is $\sim 10^{-25}$.  
Note that for PBH's to form via the necklace-specific process outlined above \emph{and} to survive to the present day, thus contributing to the 
current mass density of the universe, we require
%(4.58)
\begin{equation} \label{eq:formation_bound}
M_T(t_i) \geq M_{*}
\end{equation} 
for at least some range of $t_i$ within $t_{PBH}^{-} \leq t_i \leq t_{PBH}^{+}$. There are then three possible scenarios which we can consider.
\begin{itemize}
\item $M_* \leq \frac{\eta_s^{-1}}{2\mathcal{G}}$
If this condition is satisfied for all $t_i$ in which black holes are produced, we can integrate the first term 
in (\ref{eq:omega_eqn}) over the range 
$t_{PBH}^{-} \leq t_i \leq t_{PBH}^{+}$. 
\item  $\frac{\eta_s^{-1}}{2\mathcal{G}} < M_* < M_T^{(max)}$
With $M_{*}$ in this range, the expression above will automatically be satisfied for times between $t_*^- \leq  t_i \leq t_*^+$, which may be 
obtained by solving (\ref{eq:formation_bound}).
\item  $M_* > M_T^{(max)}$
All PBH's formed by this process will evaporate long before the present epoch. 
\end{itemize}
In reality the first of these scenarios will not occur if $\eta_s \sim \sqrt{\alpha'}^{-1} \ (\equiv M_{Pl})$ \emph{unless} $R$ is hierarchically 
larger than the fundamental string scale. 
This is theoretically compatible with the bound (\ref{eq:primeconstraint}) for small enough values of the warp factor, 
though we need not assume that the bound is close to saturation. 
Working in Planck units and using the values for $\mathcal{G}$ (\ref{eq:G_definition}) and $\eta_s$, it is possible to show 
that $M_* > \eta_s^{-1}/2\mathcal{G}$ so we need only 'tune' the values of our model parameters so that $M_T(t_i) > M_*$ for at least
 \emph{some} range of $t_i$. 
It turns out that the most important parameters for ensuring this condition is met are the warp factor $a_0$ 
and the world-sheet flux momentum $\Pi$. There is in fact a region (if somewhat restrictive) of the whole parameter space of $a_0, R, \Pi$ for which this
condition is met which we shall discuss later. But it seems that one is either required to have small values of $a_0$ and/or large values of 
$\Pi$. That large $\Pi$ should help ensure PBH production is understandable since the pre-factor $\sqrt{1 + \frac{\Pi^2}{T_1^2 \lambda^2}}$ simply multiplies the 
expression for $M_T(t_i)$, whereas increasing world-sheet flux does not affect the value of $M_*$. 
However the relation between $M_T(t_i)$ and $a_0$ is more complicated as this factor appears in a complex way inside the Elliptic function.

%However, we note that 'typical' values of $a_0 \sim 10^{-11}$ and $\Pi \sim 10^{12}$ are needed to ensure PBH production.  
%We must therefore 'tune' (though, as we shall see, not \emph{fine} tune) the values of warp factor $a_0$ and the world-sheet flux momentum $\Pi$ in order to ensure that $M_T(t_i) \geq \eta_s$ for at least \emph{some} range of $t_i$. That large $\Pi$ helps to fulfill either of the second or third conditions above is easy to see as this affects the

For the sake of completeness we calculate $t_*^{\pm}(M_*)$ explicitly by setting $M_T(t_i) = M_*$.
We then use the early time expansion of $M_T(t_i)$, given previously, to estimate $t_*^{-}$ - and the late time approximation to $M_T(t_i)$ 
to obtain $t_*^{+}$. The result is
%(4.59)
\begin{eqnarray} \label{eq:tstarpm}
t_*^{-}(M_*) &\sim &  \frac{M_* R}{ 4 T_1 a_0 \alpha^2 T_1\sqrt{1+\frac{\Pi^2}{T_1^2 \lambda^2}}}\alpha'^{-\frac{1}{2}} \nonumber \\
t_*^{+}(M_*) &\sim & \frac{\pi \alpha^2 T_1 \sqrt{1+\frac{\Pi^2}{T_1^2 \lambda^2}}}{2 a_0^3 M_*R}\alpha'^{\frac{3}{2}}
\end{eqnarray}
Although one could also obtain $t_*^{+}$ by equating $M_{Tapprox}(t_i)$ with $M_*$, the resulting expression for $t_*^{+}$ is rather unwieldy and does 
not significantly differ in its value from the approximate form given here.
The evaluation of $\Omega_{PBH}(t_0)$ (\ref{eq:omega_eqn}) between the limits $t_*^{-}$-$t_*^{+}$ 
(given by (\ref{eq:tstarpm}) above) is then obtained by integrating between $t_*^{-}$ and $t_E$ using 
the early time approximation (\ref{eq:t_E}) and between $t_E$ and $t_*^+$ using the numerical fit.
The full expression is then well approximated by the following:
%\newpage
%\begin{eqnarray} \label{eq:Omega_FULL}
%\Omega(t_0) &=& \frac{g\sqrt{\alpha}\sqrt{\Pi^2 + \lambda^2 T_1^2}}{\pi a_0 R^7 t_0^{3/2} \rho_{crit}}\left\{-4AC^{\frac{1}{4}}(F-1)^{-\frac{1}{4}}\left(\frac{F+5}{(F-1)^{\frac{1}{4}}}-3\times 2^{\frac{3}{4}}_{2}F_{1}\left(\frac{1}{4},\frac{1}{4};\frac{5}{4};\frac{1}{2}(1-F)\right)\right) \cr
%&-& 8\left(2\sqrt{\frac{32}{\pi}}(F^2 -1)^{-\frac{1}{4}}-\frac{1}{384}(log(256)-3)(F^2 -1)^{\frac{3}{4}}+\sqrt{2\pi}(1+log(16)))(F^2 -1)^{\frac{1}{4}} \cr
%&-& \pi log\left(\frac{1}{8a_0}(F^2 -1)^{\frac{1}{2}}\right)\right) - 8\pi log\left(\frac{1}{2a_0^2} - 19.312\right)\right\}
%\end{eqnarray}
\begin{eqnarray} \label{eq:Omega_FULL}
&&\Omega(t_0) \sim \nonumber\cr
&&\nonumber\cr
 &&\frac{10^{-25}g\sqrt{\alpha}\sqrt{\Pi^2 + \lambda^2 T_1^2}}{\pi a_0 R^7 t_0^{3/2} \rho_{crit}} 
 [ -4AC^{\frac{1}{4}}(F-1)^{-\frac{1}{4}} \left\lbrace \frac{F+5}{(F-1)^{\frac{1}{4}} }-3\times 2^{\frac{3}{4}}
 \, {_{2}F_{1}} \left(\frac{1}{4},\frac{1}{4};\frac{5}{4};\frac{1}{2}(1-F) \right) \right\rbrace \nonumber \cr
&&\nonumber\cr
&&\nonumber\cr
&&-( 2{\left(\frac{32}{\pi}\right)}^{\frac{1}{2}}(F^2 -1)^{-\frac{1}{4}}-\frac{1}{384}(\rm{Log}(256)-
3)(F^2 -1)^{\frac{3}{4}}+\sqrt{2\pi}(1+\rm{Log}(16))(F^2 -1)^{\frac{1}{4}} \nonumber \cr
&&\nonumber\cr
&&\nonumber\cr
&&-\pi \rm{Log} \left( \frac{1}{8a_0}(F^2 -1)^{\frac{1}{2}} \right) - 8\pi \rm{Log} \left( \frac{1}{2a_0^2} \right) -19.312 ) ] 
\end{eqnarray}
where $_{2}F_{1}$ is the usual Hypergeometric function and we have defined
$F=F(a_0,R,M_*,\alpha,\lambda T_1, \Pi)$ to be;
\begin{equation} \label{eq:F}
F = \frac{16a_0^2 M_*^2 R^2}{\alpha^4 (\Pi^2 + \lambda^2 T_1^2)}.
\end{equation}
As noted above, the fulfillment of the condition $M_T(t_i) \geq M_*$ requires small values of the warp 
factors and/or large values of the flux parameter $\Pi$ - though the exact relationship between 
$M_T(t_i)$ and $a_0$ is complex. 
We must also ensure, for consistency, that $t_*^{-}(M_*) < t_*^{+}(M_*)$  (where these values are given by the expression (\ref{eq:tstarpm})). 
%Although it is theoretically possible to determine the exact region of
%parameter space in which all these conditions are fulfilled we here content 
%ourselve's with stating that 'typical' values of $a_0 \sim 10^{-11}$ and $\Pi \sim 10^{12}$
%are necessary to ensure PBH production. 
%With this in mind we see that the potential value of 
%the $\Omega(t_0)$ integral \textcolor{red}{ref formula} for $R \sim \sqrt{\alpha'}$ 
%is \emph{huge} ($\sim 10^{3} >>1$). 
Exactly how small $a_0$ is required to be in order to fulfill both of these requirements depends on how large we take the $\Pi$ to be. 
We see from the explicit expression for $\Pi$ given below that, in principle, $0 \leq \Pi \leq \infty$ 
for flux $\lambda F_{0 \sigma}$ in the allowable range $0 \leq \lambda^2 F_{0 \sigma}^2 \leq a_0^2(a_0^2 r^2 + R^2 s'^2)$. 
\begin{equation} \label{eq:Pi}
\Pi = \frac{2T_1 \lambda^2 F_{0\sigma}^2}{a_0^2\sqrt{r^2 + a_0^{-2}R^2s'^2 - a_0^{-4}\lambda^2 F_{0\sigma}^2}}
\end{equation}
As causality requires that $F_{0\sigma}^2 < a_0^4 \lambda^{-2} (r^2 + a_0^{-2}R^2s'^2)$ we may assume,
\begin{equation}
F_{0\sigma}^2 = \beta a_0^4 \lambda^{-2} (r^2 + a_0^{-2}R^2s'^2)
\end{equation}
where $0 \leq \beta < 1$. Equation (\ref{eq:Pi}) may then be re-written in terms of $\beta$ so that,
\begin{equation}
\Pi = \frac{2T_1\sqrt{\beta}\lambda}{\sqrt{1-\beta}}
\end{equation}
from which it is clear that $\beta = 0$ corresponds to $\Pi = 0$ and that $\Pi \rightarrow \infty$ as $\beta \rightarrow 1$.
However the bound for PBH formation is not the only condition we must consider as the predictions of our model must also be 
consistent with observational bounds. Current observational constraints on the energy density of PBH's come from the EGRET experiment, 
which measures the extra-galactic gamma ray flux at 100MeV \cite{Hartman:1999fc}. 
By calculating the expected contribution to this flux from black holes expiring at the present epoch \cite{Wichoski:1998ev, MacGibbon:1991}
(see also \cite{Lehouqc:2009} for bounds derived for the standard PBH spectrum using the latest data)
they were able to show that the current density of PBH's formed via the collapse of cosmic strings is bounded by
\begin{equation}
\Omega_{PBH}(t_0) < 10^{-9}.
\end{equation}
Note however that this bound is based on the prediction that the PBH mass spectrum follows the profile 
predicted by the standard Hawking collapse process, such that $M_{PBH}(t) \sim t$ and the number density per mass interval is given by (\ref{eq:n_PBH}).
Technically one should recalculate this bound using the spectrum predicted by the necklace-specific collapse channel in order to 
place bounds on the model parameters from experimental data. We hope to supply a re-calculated bound on $\Omega_{PBH}(t_0)$ at a later date, 
but content ourselves for the time being with $\Omega_{PBH}(t_0) < 10^{-9}$ as a "ball park" figure with which to proceed.  

With this in mind we see that not \emph{all} the region of parameter space allowing PBH production is compatible with observation. 
For example, 'typical' values of $a_0 \sim 10^{-11}$ and $\Pi \sim 10^{12}$ are sufficient to ensure that $M_T(t_i)>M_*$ for at least 
some $t_i$ but the resulting value of 
the $\Omega_{PBH}(t_0)$ integral (\ref{eq:Omega_FULL})is \emph{huge} ($\Omega_{PBH}(t_0) \sim 10^{3} >>1$) if $R$ is comparable 
to the string scale.
However due to the large $R$-dependence in the 
denominator caused by the extra-dimensional contribution to $\nu_r$ we (happily) see 
that typical values of $R \sim 10^2$ are sufficient to bring $\Omega_{PBH}(t_0)$ within the 
observable bound $\Omega_{PBH}(t_0) < 10^{-9}$. Thus the dimensions of the compact space 
need not be hierarchically larger than the string scale in order for the
predictions of our model to be consistent with observational constraints.
	
To systematically explore the values of $a_0,R $ and flux $ \Pi$ which are consistent with all the constraints above, we can  
express the value of the peak of the necklace mass function, $M_{Tapprox}$ given earlier in 
(\ref{eq:MT_A_C}), in terms of $M_*$;
\begin{equation}\label{eq:kappa}
{M_T}^{peak} \sim 5.47 \ T_1  \sqrt{1+\frac{\Pi^2}{T_1^2 \lambda^2}} \frac{\alpha^2 \alpha'^{\frac{3}{2}}}{8a_0 R} = 
\kappa M*
\end{equation}
where $\kappa >1$ is a parameter that for given values of $(a_0,\alpha, R, \Pi, T_1, \alpha')$ 
express how far the peak value of $M_T$ is above the mass scale $M_*$.
Using this definition we can expresses the quantity $\sqrt{1+\frac{\Pi^2}{T_1^2 \lambda^2}}$ in terms of $(a_0,\alpha, R, \kappa, T_1, \alpha') $
and the above terms for $ t_*^{\pm}$ simplify considerably
\begin{equation}\label{eq:tstarsimp}
t_*^{-} = \frac{0.171 {\alpha'}^{-1/2}}{a_0^2 T_1 \kappa},  \qquad  t_*^{+} = \frac{2.297T_1 \kappa {\alpha'}^{3/2}}{a_0^2} 
\end{equation}
It is interesting to note that both $ t_*^{-}$ and $ t_*^{+}$ depend on the warp factor as $a_0^{-2}$.
It may be thought that for $\kappa =1$,  the times $t_*^{-}$ and $t_*^{+}$ should approach one another since
the definition of $t_*$ requires values of $t_i$ for which $M_T = M_*$. However recall we have assumed the early time approximation for $M_T$ in determining 
$t_*^{-}$, and not the 'late-time' approximation $M_{Tapprox}$. This assumption then requires that we take $\kappa > 1$.
We could of course consider situations where $\kappa$ is closer to unity, but then one has to use $M_{Tapprox}$ in 
the determination of both $ t_*^{-}$ and $ t_*^{+}$ - which is technically complicated.

Now evaluating (\ref{eq:Omega_FULL}) using the above, we obtain an expression that depends on the parameters
$ (a_0,\alpha, R, \Pi, T_1, \alpha', t_0, \rho_{crit})$. In Figure 8 we present a contour plot showing the values of $R, \kappa$ 
that are consistent with the condition  $\Omega_{PBH}(t_0) < 10^{-9}$. 
In this plot we have taken, as a typical value, $\alpha=0.8$ and have set $T_1=\alpha'^{-1}=1$ and 
input the standard values of $\rho_{crit}$, $t_0$ and $M_*$. The resulting bounds on $\lbrace R, \kappa \rbrace$ are remarkably 
insensitive to the actual value of $a_0$ because the latter only appears in $\Omega_{PBH}(t_0)$ via logarithmic factors. 
 
For given (allowed) values of $\lbrace R,\kappa \rbrace$ we can then deduce the value for the flux parameter $\Pi$ via (\ref{eq:kappa}), 
which is of course sensitive to the value of $a_0$.
%Figures 8
\begin{figure}[htp]
\begin{center}
\includegraphics[width=0.9\textwidth]{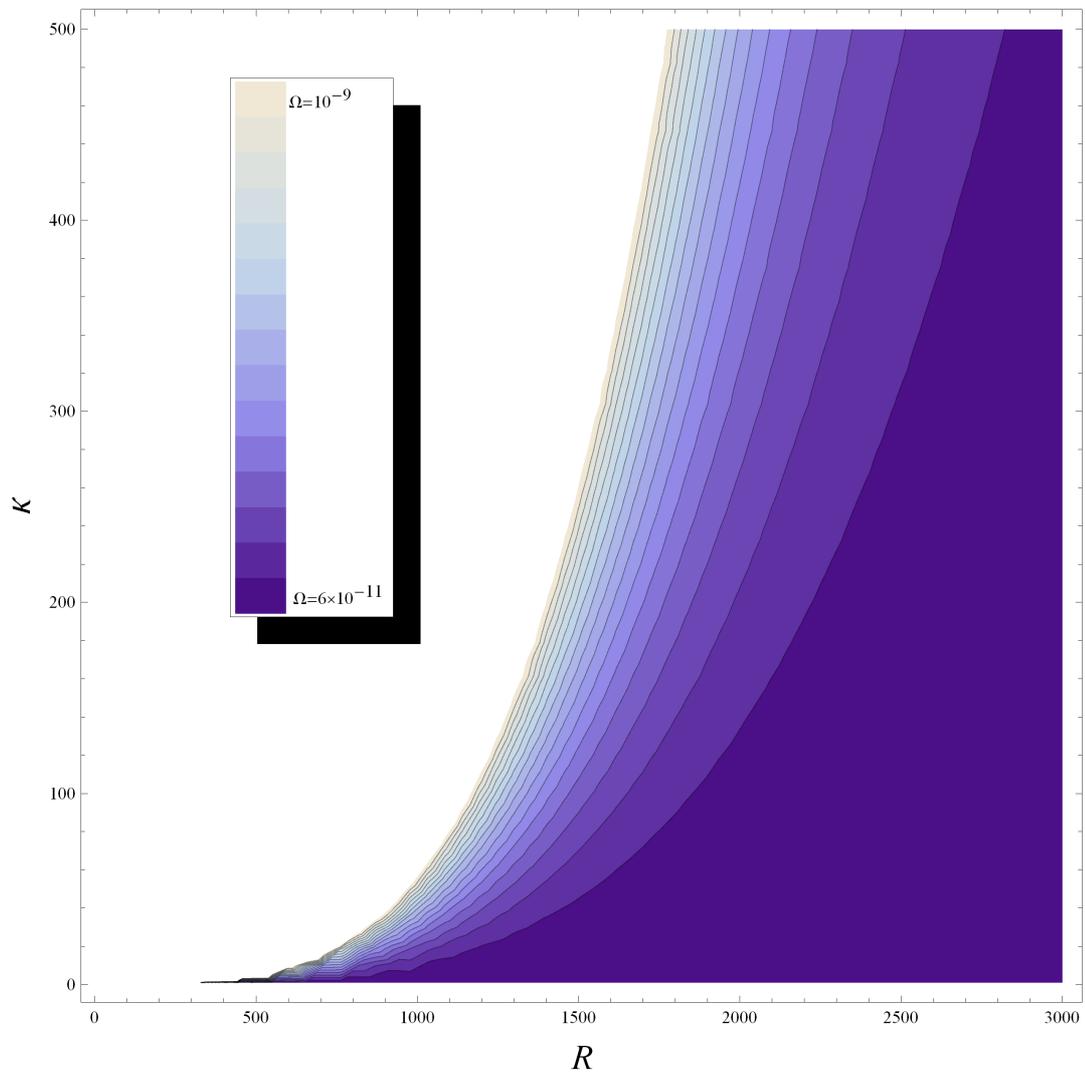}
\caption{Contour plot showing the allowed regions of $\lbrace R,\kappa \rbrace $ consistent with the bound $\Omega(t_0) <10^{-9} $. 
Only a few contours are plotted. We have chosen $a_0=0.1, T_1 = 1 = \alpha'$ and $\alpha=0.8$}
\end{center}
\end{figure}
To visualize the allowed values of the flux parameter $\Pi$ corresponding to those of $a_0, R, \kappa$ - we use a 3D-contour plot shown in Figure 9.
We illustrate six typical surfaces in the $\lbrace \rm{Log} [a_0], R , \kappa \rbrace$ plane corresponding to values of 
$\Pi = 10^{12},10^{10},10^8,10^6,10^4,10^2$ (from front to back). These surfaces also take into account the 
observational allowed values of $\lbrace R, \kappa \rbrace$ illustrated in Figure 8. 
%Figure 9
\begin{figure}[htp]
\begin{center}
\includegraphics[width=0.5\textwidth]{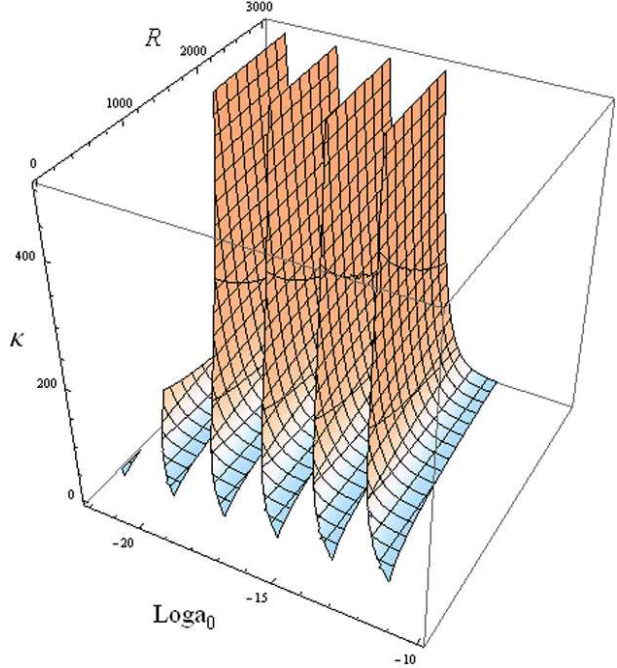}
\caption{ 3D Contour plots showing values of the flux parameter $\Pi$ in the  ($\rm{Log}[a_0], R,\kappa $) plane.
Six surfaces are shown corresponding (from front to rear) to the values of $\Pi= 10^{12},10^{10},10^8,10^6,10^4,10^2$. 
These plots take into account the observational bounds on ${R,\kappa}$ shown in Figure 8. The remaining choice of
parameters are as in Figure 8.}
\end{center}
\end{figure}

\subsection{Quasi-stable necklaces}
Let us  now drop the assumption that the loops retain their necklace structure indefinitely. 
As outlined earlier, it is reasonable to expect that wrappings in the flat $\phi$-direction may unwind over time leading to a "flattening" of the $\theta$-direction. 
This in turn flattens the $\psi$ direction and, so on until the whole necklace structure unravels.

Assuming that the motion of the string in the $\phi$-direction remains random \emph{after} the formation of a loop, 
and that the motion is also random in any newly flattened direction, we should expect the characteristic lifetime of the 
beads in any necklace to be comparable to the formation time, $t_i$. This is because the average warped time taken for a single $\phi$-winding 
to contract to a point is $a_0 t_1 \sim \frac{R^2}{a_0^{-1}\sqrt{\alpha'}}$ - where this is the time associated with a single winding
\footnote{Here we have used the fact that the radius of the winding must contract by a distance $\sim R$. 
Assuming a step length of $\epsilon_l \sim a_0^{-1}\sqrt{\alpha'}$ for the change in the radial coordinate 
this requires a total displacement of $\sim \frac{R}{a_0^{-1}\sqrt{\alpha'}}$ steps. 
This in turn requires on average $\sim R^2/{a_0^{-2} \alpha'}$ random steps which takes (a warped) time 
$a_0 t \sim \frac{R^2}{a_0^{-2}\alpha'} \times a_0^{-1} \sqrt{\alpha'} = \frac{R^2}{a_0^{-1}\sqrt{\alpha'}}$.}. 
The warped time taken for $n(t_i)$ such windings to contract is therefore
\begin{equation}
a_0 t_{n(t_i)} \sim \frac{n^2(t_i) R^2}{a_0^{-1} \sqrt{\alpha'}}
\end{equation} 
As we are only considering that part of the string which forms the extra-dimensional windings 
we may compare the expression above with (\ref{eq:n}) in which $\omega_l$ and $\alpha$ have been set to unity. We then see that,
%\label{eq:t_ti}
\begin{equation} \label{eq:t_ti}
t_{n(t_i)} \sim t_i
\end{equation}
This implies that the number of necklaces originally formed at $t_i$ which survive for a time $\Delta t >> t_i$ \emph{after} formation will be negligible, 
with most having become standard string loops. However it is difficult to estimate the fraction of necklaces surviving for an arbitrary 
length of time $\Delta t$ using such generic arguments, though this is exactly what we must calculate to determine the true contribution 
of necklace collapse to the PBH mass spectrum. 
In particular we must calculate the fraction of loops which retain their necklace 
structure for an interval $\Delta t \sim \frac{\alpha}{\Gamma \mathcal{G} T_1}t_i $.  To do this we consider the probability distribution 
which describes fluctuations of the radial coordinate. For a random walk this is simply a Gaussian distribution with mean $\mu = 0$ and variance, 
\begin{equation}
\sigma^2(\Delta t) \sim a_0^{-1}\frac{\Delta t}{\delta t} \epsilon_l^2 \sim  a_0^{-2}\sqrt{\alpha'} \Delta t
\end{equation}
where $\delta t \sim a_0^{-1}\sqrt{\alpha'}$ is the time interval between steps, $\epsilon_l \sim a_0^{-1}\sqrt{\alpha'}$ 
is the step length and $\Delta t$ is the total (un-warped) time elapsed. 
The total probability density function $\Phi(t,\sigma(\Delta t))$ is given below.
\begin{eqnarray}
\Phi(t,\sigma(\Delta t)) &\sim& \frac{1}{\sqrt{2\pi} \sigma(\Delta t)} e^{-\frac{1}{2} \frac{t^2}{\sigma^2(\Delta t)}}  \nonumber \\
&\sim& \frac{a_0}{\sqrt{2 \pi} (\sqrt{\alpha'} \Delta t)^{\frac{1}{2}}} \exp \left(-\frac{a_0^2 t^2}{2 \sqrt{\alpha'} \Delta t} \right).
\end{eqnarray}
Since a loop forming at time $t_i$ has $n(t_i)$ windings to lose, the radial coordinate must "travel" a distance 
approximately equal to $n(t_i)R$ in order for the loop to lose its necklace structure. 
The fraction of necklaces $f(\Delta t)$ which survive for an interval $\Delta t$ after $t_i$ is then given by the integral of the above 
expression between $t \sim \pm a_0^{-1}n(t_i)R \sim \pm a_0^{-1} \left(\sqrt{\alpha' t_i}\right)^{\frac{1}{2}} \sim \pm \sigma(t_i)$
which can be well approximated using the error function
\begin{eqnarray}
f(\Delta t) &\sim& Erf \left(\frac{\sigma(t_i)}{\sqrt{2}\sigma(\Delta t)}\right) \sim Erf \left(\frac{1}{\sqrt{2}}\frac{a_0^{-1}\left(\sqrt{\alpha' t_i}\right)^{\frac{1}{2}}}{a_0^{-1}\left(\sqrt{\alpha' \Delta t}\right)^{\frac{1}{2}}}\right) \nonumber\\
&\sim& Erf \left(\sqrt{\frac{t_i}{2\Delta t}}\right).
\end{eqnarray}
Thus when $\Delta t \sim t_i$ the fraction of loops which have retained their extra-dimensional windings is approximately
\begin{equation}
f(t_i) \sim Erf\left(\frac{1}{\sqrt{2}}\right) \sim 0.68
\end{equation}
which matches well with our result (\ref{eq:t_ti})
\footnote{Note that $f(\Delta t) < 0.7$ for all $\Delta t > t_i$. 
In fact $Erf(1/2) \sim 1/2$, hence the \emph{majority} of loops ($f(\Delta t) > \frac{1}{2}$) will have lost all their windings by $\Delta t \sim 2t_i$. 
This helps to quantify our earlier result (\ref{eq:t_ti})  more precisely.}.
The fraction of necklaces which survive until they reach the minimum radius is,  
\begin{equation}
f\left(\Delta t \sim \frac{\alpha}{\Gamma \mathcal{G} T_1} t_i\right) 
\sim  Erf \left(\frac{1}{\sqrt{2}}\left(\frac{\alpha}{\Gamma \mathcal{G} T_1}\right)^{-\frac{1}{2}}\right)
\end{equation}  
Strictly speaking therefore the integral (\ref{eq:omega_eqn}) should be multiplied by the additional numeric factor 
above to account for the the loss of necklaces which un-ravel before reaching the point at which they can undergo gravitational collapse.

We also note that $f\left(\Delta t \sim \frac{\alpha}{\Gamma \mathcal{G} T_1} t_i\right) \leq 10^{-20}$ when,
\begin{equation}
\alpha \ge 1.591 \times 10^{39} \Gamma \mathcal{G} T_1
\end{equation}  
and thus the necklace-specific channel becomes comparable (or sub-dominant) to the standard Hawking process. Practically however the values of $\alpha$, $\Gamma$, $\mathcal{G}$ and $T_1$ here are such that this is unlikely to occur without significant fine tuning.
%%%%%%%%%%%%%%%%%%%%%%%%%%%%%%%%%%%%%%%%%%%%%%%
\subsection{Comparison with the standard string-monopole network}
The dynamics of string-monopole network evolution depends crucially upon the ratio of the monopole energy density 
to the string tension \cite{Berezinsky:1997td, Matsuda:2006ju}. This dimensionless parameter, denoted $\mathcal{R}$ in the literature is given by,
\begin{equation}\label{eq:Rdefinition}
\mathcal{R} = \frac{m}{\mu d}
\end{equation}
where $m$ is the the monopole mass, $\mu$ is the tension and $d$ is the distance between monopoles. 
For $\mathcal{R} << 1$ the network behaves like an ordinary string network - reaching a scaling solution at late times, 
but for $\mathcal{R} >> 1$ the mass of the beads dominates the dynamics of the evolution. 
In our case these energy scales are time dependent and the corresponding parameter is given by
\begin{equation}\label{eq:Rdefinition_us}
\mathcal{R}(t_i) = \frac{M_b(t_i)}{T_1 d(t_i)}
\end{equation}
which scales as $\mathcal{R} \sim t_i^{-2}$ for $t_i >> \frac{2\sqrt{\alpha'}}{a_0^2 \alpha}$ 
and $\mathcal{R} \sim t_i^{-\frac{1}{2}}$ for $t_i << \frac{2\sqrt{\alpha'}}{a_0^2 \alpha}$. Thus we see that as $t_i \to \infty$ we find $\mathcal{R} \to 0$ and vice-versa.
Our results are therefore consistent with the standard analysis in that the necklaces behave like an ordinary string network for $\mathcal{R} << 1$ 
but like a string-monopole network when $\mathcal{R} >> 1$ since the mass of the beads becomes significant. 

However the dynamics of $\mathcal{R}$ differ profoundly in our model. 
This indicates that the effect of beads formed from extra-dimensional windings is different to the effect created by monopoles 
\footnote{For example monopoles formed after a separate phase transition at some temperature $T_M > T_s$.} with regard to network evolution. 
To understand this in more detail let us consider the latter.
The standard equation for the evolution of $\mathcal{R}$ is \cite{Matsuda:2005fb},
\begin{equation}\label{eq:Rdynamics}
\frac{\dot{\mathcal{R}}}{\mathcal{R}} = -\kappa_s t^{-1} + \kappa_g t^{-1}
\end{equation}
where the first term on the right hand side describes the stretching of the string due to the cosmic expansion 
and the the second term describes the contraction due to emission of gravitational radiation.  
It is a reasonable to assume that
\begin{equation}
\kappa = \kappa_g - \kappa_s > 0  
\end{equation}
which allows us to solve (\ref{eq:Rdynamics}), up to some constant of integration,
\begin{equation}
\mathcal{R} \sim t^{\kappa} .
\end{equation}
Using (\ref{eq:Rdefinition}) the evolution of the inter-monopole distance then scales like
\begin{equation}
d \sim \mathcal{R}^{-1} \sim t^{-\kappa}. 
\end{equation}
This was the crucial assumption Matsuda used to identify $d(t)$ with the step length of the random walk $\chi(t)$ \cite{Matsuda:2005fb} of the monopole along the string.
From this point, the estimated initial number of beads per loop is then given by
\begin{equation}
n(t_i) \sim \frac{r(t_i)}{d(t_s) \times \left(\frac{t}{t_s}\right)^{k-1}} 
\end{equation}
where $d(t_s) \sim (t_M t_s)^{\frac{1}{2}}$ is the initial bead spacing in the \emph{network} at the time of string formation $t_s$ 
(where here $t_s \sim \sqrt{\alpha'}$) and $k=0$ corresponds to the (natural) $\kappa = 1$ solution of (\ref{eq:Rdynamics}).

Assuming that $\sim \sqrt{n}$ beads would survive without annihilation, we find that $m_{coil} \propto \sqrt{n}$ and hence,
\begin{equation} \label{eq:coil}
m_{coil} \sim t_i^{\frac{2-k}{2}} 
\end{equation}  
which favours black hole production at late times vs dark matter production for small $t_i$. This is the standard result assuming that necklaces which reach their minimum radius but which have insufficient mass to undergo collapse to form PBH's can only interact with other matter gravitationally, leading to the production of Planck scale dark matter relics. In Matsuda's original scenario however, based on the assumptions above, large numbers of Planck scale dark matter 'particles' are formed at early times with smaller numbers of increasingly massive black holes formed at later times. In our model this process is essentially reversed with a window of PBH production in the early universe (though potentially this window may be quite large for small $a_0$ so the last epoch of PBH formation is not necessarily 'early', even on cosmological time scales) when $M_T(t_i) > M_*$ with dark matter forming well into the scaling regime when $M_T(t_i) < M_*$.

Physically the above equations describe the shrinking of the string sections connecting neighbouring monopoles. 
Effectively the contraction of the string is able to pull these monopoles through the horizon at an ever increasing rate 
\footnote{Alternatively this can be viewed as the expanding horizon uncovering monopoles, separated by 
increasingly short distances, the distances having been shortened by the contraction of the string.}. 
Hence we find that the early and late time limits of $\mathcal{R}$ as defined above, are the opposite of those obtained from (\ref{eq:Rdefinition_us}), 
namely that $\mathcal{R} \to \infty$ as time increases, whilst tending to zero as $t_i \to 0$.

Similarly the natural solution, $\kappa = \kappa_g - \kappa_s \sim 1$ obtained from an order of magnitude estimate, 
gives $d \sim t^{-1}$ in the standard case as opposed to $d \sim t$ in our case. 
In yet another respect therefore we have obtained the opposite results to the standard analysis. 

However it is not immediately obvious that what applies to monopoles formed in a separate phase transition also applies to beads 
formed by extra-dimensional windings. Although a contracting string may pull ordinary monopoles through the horizon at an ever increasing rate, 
a winding may not cross the horizon ready-made and concentrated at a point from a four-dimensional perspective. 
This is because insufficient time will have elapsed to establish correlations in the compact dimensions. Put another way, if the horizon advances by a distance $a_0 c \delta t$, the end point of the string formerly localised at the horizon, 
can move by, at most, a distance $c \delta t$ from its original position in the compact space. 
Causality therefore places a limit on the rate at which new windings can enter the horizon, 
whereas no such limit exist for ordinary monopoles 
\footnote{An alternative way to view this result is to consider windings as correlations in the compact space. 
All windings must therefore form within the horizon to preserve the causal structure.}.

%In our case it is quite impossible for even a single bead, corresponding to a winding of $\sim \frac{1}{2}R$, to come through 
%the horizon within a time $\delta t \sim \sqrt{\alpha'}$ if $R > 2\sqrt{\alpha'}$. 
%Note that this condition corresponds to equation(5.16) which is itself a bound derived from the causal structure. 
%If instead we take the average effective radius $<R>$ of a winding on the $S^3$, we find  $<R> \sim \frac{R}{\sqrt{2}}$ then the constraint on $R$ 
%changes to become $R > 2\sqrt{2 \alpha'}$.
In fact the considerations above help to explain why the number of windings $n(t_i)$ is proportional to the ratio $a_0/R$ not simply to $R^{-1}$ in the warped throat model. 
%At first sight we might have expected that a smaller radius $R$ would promote winding formation at earlier times, though (5.16) and (5.36) 
%suggest the opposite 
%\footnote{For example, if $R$ lies in the range $2\sqrt{\alpha'} < R < 2\sqrt{2\alpha'}$, beads will form in the scaling epoch but not during the earlier damping regime.}. 
Although larger $R$ results in a slower rate of winding formation (as we would expect) this effect is, at least potentially, dwarfed by the effect of the warp factor $a_0$ (c.f. (\ref{eq:n_squared})) which is related to $R$ via the deformation parameter, $a_0^2 \sim \tilde\epsilon^{\frac{4}{3}}R^{-2}$ 
This is because $a_0$ limits the increase of the horizon distance in the infinite directions (but not in the compact space) via
\begin{equation}
d_H(\delta t) = a_0 c \delta t
\end{equation}   
As the \emph{resultant} velocity of the end point of the string is $c=1$ smaller $a_0$ results in a greater velocity in the compact spaces and hence a larger limiting value for $n(t_i)$.

In general however, windings may enter the horizon more rapidly the smaller their effective radii. 
In fact this at least partially accounts for the falling bead mass at late times; since
 $t_i \rightarrow \infty$ we expect that $\omega_l \rightarrow 0$ for a random walk in the extra dimensions. This could be achieved through either;
\begin{itemize}
\item Falling number density (per unit distance along the string) at late times - $n \sim d^{-1}$
\item Windings wrapping ever smaller effective radii in the $S^3$
\end{itemize}
The fact that $d^{-1} \sim t_i^{-1}$ at late times highlights the first effect and $M_b \sim t_i^{-1}$ indicates the second. 
In practice however we would expect both effects to play some role. 

The question then arises, does the shrinking bead mass have a counterpart in field theory? 
We may also consider whether the arguments for the limitation of windings entering the horizon hold true for 
field-theoretic strings in the dual non-Abelian Higgs models.The present authors hope to address this question in a future publication.

%\textcolor{blue}{In fact there is no reason why all the arguments for the limitation of windings entering the horizon cannot apply to non-abelian field-theoretic strings. In this case the "windings" are not around the compact directions but around the moduli space of the field Lagrangian. The 4D "distance" between windings is then the distance over which the field configuration interpolates between degenerate values in the vacuum manifold by rotating the phase at each point in the vortex. The rate at which such windings enter the horizon is then limited by the rate at which the vortex can rotate. This of course is itself limited by causality.} 
%\textcolor{blue}{In the field theory description the beads form when the field interpolates between degenerate moduli over real values (i.e. over the hill in the potential term of the Lagrangian) thus forming "kinks" along the string which are analagous to domain wall solutions. - This is what I THINK Matsuda was talking about, can someone check this? - The shrinking bead mass then corresponds to the field moving "over" the hill at distances further and further from the centre. Hence there are in principle no discrepancies between the dual higher dimensional F-string and non-albelian Higgs field descriptions, as expected.}  
Finally we note that the predictions of our model (specifically that PBH production is favoured at early times with dark matter 
production at later epochs, in contrast to previous predictions) are highly dependent on the assumption of a random walk regime which 
produces the extra-dimensional windings. It is not immediately clear how this behaviour might be modified by the introduction 
of a velocity correlations regime, though it is possible that the results may be more in line with those of previous studies. 
This is a question the authors hope to pursue at a later date. 
It is also highly dependent upon the assumption that only beads formed from \emph{net} windings, not from bead-anti-bead ($b-\bar{b}$) pair formation, 
contribute to the mass of PBH's/DM relics. As we have seen, assuming that $\sim \sqrt{n}$ $b$/$\bar{b}$'s survive until the string reaches it's minimum radius 
(where $n$ here must be taken to include \emph{all} beads/anti-beads created from random steps along the string, not simply those formed by net displacement 
in the compact space) leads to (\ref{eq:coil}) and to qualitative behaviour which is the very opposite to that obtained in our model. 
However as stated previously in the footnote on page 15, we believe that such an argument is valid only for a \emph{static} string 
and that in a contracting loop all $b$'s will eventually collide with their $\bar{b}$ counterparts before the minimum radius is reached. 
In particular, in our model we may expect the initial $b-\bar{b}$ spacing to be approximately equal to the step length $\epsilon_l \sim a_0^{-1} \alpha \sqrt{\alpha'}$ 
which is most certainly greater that the minimum radius $\delta_l \sim \eta_s^{-1} \sim \sqrt{\alpha'}$ for all reasonable values of $\alpha$ and $a_0$.
%%%%%%%%%%%%%%%%%%%%%%%%%%%%%%%%%%%%%%%%%%%%%%%%%%%%%%%%%%%%
\section{Discussion}
In this note we have investigated a simple cosmic string model using several key ideas from string theory. Our starting point was the assumption that
there is some initial stage of inflation that leads to the creation of trapped strings (along the lines of \cite{Sarangi:2002yt}) 
that have non-zero windings in the internal space.
We argued that the string feels a lifting potential due to the non-trivial geometry of the background, forming a `kinky cycloop'.
By estimating the mass of such a string configuration we were able to show that 
formation at late times reduces to a scaling solution, which one may hope to identify with a relic
dark matter phase. We also note in passing that, in principle, we require dark matter production from necklaces between $t_{PBH}^{+}$ and $t_0$ not to over-close the universe. However, as the present day dark matter density bound is approximately $\Omega_{DM}(t_0) < 0.3$ and due to the $t^{-3}$ suppression of the loop production function, we find that in all but pathological cases the PBH density bound $\Omega_{PBH}(t_0) < 10^{-9}$ is by far the most stringent constraint.

%At very early times, the bead mass is explicitly time dependent resulting in a small window for PBH formation when (in the saturated limit)
%\begin{equation}
%0 \le t_i \le \frac{\sqrt{2\alpha'}}{\alpha a_0^2}
%\end{equation}
%which can be sufficiently long lived for a small enough warp factor. The warp factor itself is bounded from above in this model using causality arguments, with
%early time solutions allowing a significantly smaller warping than at late times.
In some sense this has been a bottom up approach to the problem. The warp factor is simply a parameter of the metric, depending only on the extra-dimensional 
flux and the deformation parameter of the (non-compact) supergravity background. A fully realised UV approach will precisely fix the warping as a function
of the closed string moduli VEV's and the flux, therefore leading to a more tightly constrained theory.
However we would hope that the general features of our approach will remain intact. In particular
the formation window for PBH at early times contrasts strongly with the field theory/string model proposed by Matsuda. Although the authors initially feared that result may be in some sense disappointing, due to the likelihood of PBH's formed at early times having evaporated by the present epoch, we have shown that it is possible within the model here constructed to satisfy the condition $M_T(t_i) > M_*$ \emph{and} current observational constraints. It is worth noting too at this point that, for very small $a_0$, the window of black hole formation may be sufficiently large for the assumption of a constant $M_*$ to become inaccurate. A more detailed analysis would therefore take into account the rate of black hole evaporation in the extra-dimensional background introducing a time-dependent lower bound on the PBH mass $M_*(t_i)$. An improved analysis would also consider how the abundance of PBH's formed through necklace collapse in the very early universe affects BBN and subsequent structure formation. In fact, there are \emph{many} theoretical and observational constraints which any such model must satisfy concurrently, thereby potentially allowing the Klebanov-Strassler model parameters to be bounded with even greater precision.

It is clear from our model that in order to generate any PBH's of mass $M_*$ restricts us to a small region of the $(a_0, R, \Pi)$ parameter space as is shown by 
%(\ref{eq:MT_A_C}).
in Figures 8 and 9. This is not surprising if we recall that $M_* \sim 10^{20}$ in Planck units and so there is a hierarchy issue here. 
On the other hand since the warp factor $a_0$ depends exponentially on background fluxes, obtaining small enough values to generate a hierarchically large $M_T$ is perhaps not so unnatural.

In order to satisfy the PBH observational bounds one requires larger background flux $M$ if there is larger world-volume flux $\Pi$. 
This is inherently obvious, since the additional world-volume flux can be treated as an effective mass correction to the necklace. This additional mass will back-react on the solution and the probe limit will be rendered invalid, unless the background flux is also increased to compensate. What is interesting is that, for fixed warping, there
is a large parameter space where bounds are satisfied. Another interesting feature is that, as is clear from Figure 8, the allowed values of $R$
for regions where $M_T >> M_*$ (corresponding to large $\kappa$) are still in the 'mild' hierarchy region $R \sim 10^3 l_s$. Indeed the only way to have a value of 
$R \sim l_s$ and consistent with observation requires a fine tuning where the peak of $M_T$ is taken very close to but just above $M_*$. This then reduces the 
time interval of mass $M_*$ PBH formation given by $(t_{*}^{-}- t_{*}^{+})$ which can be made small enough so that we can reduce the value of $R$ 
and still obtain $\Omega_{PBH}(t_0) < 10^{-9}$.

Our construction is very model dependent in some sense, since it is only valid for a special class of supergravity backgrounds of the type IIB string.
Another interesting class of cosmic string models arises from considering Heterotic M-theory on the orbifold $CY_3 \times S^1/\mathbb{Z}_2$. The strings in this
case can arise from wrapping membranes (or five-branes) over various cycles within the internal space \cite{Becker:2005pv, Buchbinder:2006ab}. 
Topologically stable strings are only possible for five-branes wrapping a complete four-cycle within the CY space \cite{Gwyn:2008fe}. However,
other wrappings are potentially possible if there is an uplifting potential and a similar analysis can be performed. We leave this for a future 
publication.

One other issue that remains is the validity of substituting the expression $r(t_i) = \alpha t_i$ for the variable $r$ in the bead mass terms of the EllipticE expansion and $r(t,t_i) = \alpha t_i - \Gamma T_1 \mathcal{G}(t-t_i)$ into the leading order term representing the mass of the four-dimensional sections of the string. As stated previously, this is entirely consistent with Matsuda's original assumption that the bead mass remains constant after loop production due to the "trapping" effects of the lifting potential, whereas the sections of string not contained in the extra-dimensional windings shrink through the emission of gravitational radiation. However, the re-paramaterisation invariance of the Nambu-Goto action implies that it may not be possible to talk of one 'section' of the string expanding/contracting while other sections remain fixed. By definition strings have no sub-structure which would seem to imply that they must expand or contract along their entire length. In our model this would mean that the contraction of the string due to gravitational radiation in the warped Minkowski directions in fact produces a contraction of the string along the internal direction as well. If true this implies that the expression $r(t,t_i)$ should be substituted in \emph{both} the bead-mass and four-dimensional parts of the string mass integral, thus creating an explicitly time dependent bead-mass $M_b(t,t_i)$ for $t>t_i$. Potentially this may alter the present results significantly, as we would expect the number of necklaces which retain sufficient mass in their extra-dimensional windings to produce PBH's when reaching their minimum radius to diminish sharply. This is therefore likely to lead to a much smaller amount of black hole production and greatly increased dark matter formation. It is hoped that the question of the explicit time dependence of the bead mass may be resolved conclusively by investigating field theory duals of the string necklaces discussed here. Again, the present authors hope to be able to offer some contribution in this direction in the near future.

A further aside concerns the question of binding energy between $b-b$ or $\bar{b}-\bar{b}$ pairs which coalesce. 
In field theoretic models the binding energy between true monopole or anti-monopole pairs may be calculated. 
We may suspect that, if a similar binding energy exists between beads formed from extra-dimensional windings, it's presence may alter the string dynamics 
so that the evolution of a string configuration with $n$ 'one step' beads into a single bead formed from $n$ steps will be energetically favoured. 
Once formed it is also possible that such a configuration may un-wind more rapidly than it's corresponding $n$-bead counterpart, 
leading to a necessary modification of the arguments in Section 5.3. As a first approximation we have assumed that there is \emph{no} binding energy between 
winding states. In the absence of a full BSFT picture this seems intuitively likely, given that the string must be considered as a collective phenomenon, 
which suggests that the unwinding of $n$ single steps will be just as hard as un-winding a single $n$-step wrapping. 
However it is not clear how one can really understand this problem without a more detailed analysis of the boundary theory.

Finally it is interesting to speculate further on the recent result implied by the PAMELA experiment \cite{Boezio:2008zz,Adriani:2008zr} which suggests
that there is an over-abundance of high energy positrons in the cosmic background. Rather than interpreting this as a signal for dark matter, it has
also been suggested that this is in agreement with predictions from cosmic-strings \cite{Brandenberger:2009ia}. It would certainly be interesting to
extend our analysis along those lines.
%%%%%%%%%%%%%%%%%%%%%%%%%%%%%%%%%%%%%%%%%%%%%%%%%%%%%%%%%%%%%%%%%%%%%%%%%
\begin{center}
{\bf Acknowledgments}
\end{center}
We wish to thank Levon Pogosian for a useful discussion. M.L is supported by an STFC studentship and the Queen Mary EPSTAR consortium.
This work was supported in part by NSERC of Canada. Thanks also to the referee for their clarifying remarks.

%\textcolor{blue}{The authors would also like to thank the referee for helpful and constructive comments in relation to the question of bead-anti-bead pair formation, the binding energy between bead (or anti-bead) pairs and the fine tuning of the pre-inflationary theory, as well as for the discovery of a number of typos. These comments prompted us to re-examine and clarify some important issues.} 
\begin{center}
\emph{We dedicate this work to the memory of David Birtles.}
\end{center}
%%%%%%%%%%%%%%%%%%%%%%%%%%%%%%%%%%%%%%%%%%%%%%%

\end{document}